\RequirePackage{fix-cm}
\documentclass[twocolumn]{svjour3}          
\smartqed  
\usepackage{amssymb}
\usepackage{siunitx}
\usepackage{natbib}        
\usepackage{amsmath}
\usepackage{graphicx}
\usepackage[T1]{fontenc}
\usepackage[utf8,latin1]{inputenc}
\usepackage{epsfig}
\usepackage{txfonts}
\usepackage{url}
\usepackage{color}
\usepackage{xcolor}

\usepackage{bm}
\usepackage{sgl-commands}
\usepackage[normalem]{ulem}

\journalname{my journal}

\begin{document}

\title{Strong gravitational lensing as a probe of dark matter}



\author{S.Vegetti$^{1}$, S. Birrer$^{2}$, G. Despali$^{3,4}$, C.~D. Fassnacht$^{5}$, D. Gilman$^{6}$, Y. Hezaveh$^{7,8,9,10,11}$, L. Perreault Levasseur$^{7,8,9,10,11}$, J.~P. McKean$^{12,13}$, D.~M. Powell$^{1}$, C.~M. O'Riordan$^{1}$, G. Vernardos$^{14}$}

\authorrunning{Vegetti} 

\institute{ 
    1. Max Planck Institut f\"{u}r Astrophysik, Karl-Schwarzschild-Strasse 1, D-85748 Garching bei M\"{u}nchen, Germany\\
    2. Department of Physics and Astronomy, Stony Brook University, Stony Brook, NY 11794, USA\\
    3. Dipartimento di Fisica e Astronomia "A. Righi", Alma Mater Studiorum Universit\`a di Bologna, Via Piero Gobetti 93/2, I-40129 Bologna, Italy\\
    4. Institut f\"{u}r Theoretische Astrophysik, Zentrum f\"{u}r Astronomie, Heidelberg Universit\"{a}t, Albert-Ueberle-Str. 2, 69120, Heidelberg, Germany\\
    5. Dept. of Physics and Astronomy, University of California, Davis, 1 Shields Ave. Davis, CA 95616, USA\\
    6. Department of Astronomy and Astrophysics, University of Toronto, Toronto, ON, M5S 3H4, Canada\\
    7. Department of Physics, Universit\'{e} de Montr\'{e}al, Montr\'{e}al, Canada\\
    8. Mila - Quebec Artificial Intelligence Institute, Montr\'{e}al, Canada\\
    9. Ciela - Montreal Institute for Astrophysical Data Analysis and Machine Learning, Montr\'{e}al, Canada\\
    10. Center for Computational Astrophysics, Flatiron Institute, 162 5th Avenue, 10010, New York, NY, USA\\
    11. Perimeter Institute for Theoretical Physics, Waterloo, Ontario, Canada, N2L 2Y5\\
    12. Kapteyn Astronomical Institute, University of Groningen, PO Box 800, NL-9700 AV Groningen, The Netherlands\\
    13. ASTRON, Netherlands Institute for Radio Astronomy, PO Box 2, NL-7990 AA Dwingeloo, The Netherlands\\
    14. Institute of Physics, Laboratory of Astrophysique, \'Ecole Polytechnique F\'ed\'erale de Lausanne (EPFL), Observatoire de Sauverny, 1290 Versoix, Switzerland\\
     \\
    \email{svegetti@mpa-garching.mpg.de}\\
}

\date{Received: date / Accepted: date}


\maketitle

\begin{abstract}

Dark matter structures within strong gravitational lens galaxies and along their line of sight leave a gravitational imprint on the multiple images of lensed sources. Strong gravitational lensing provides, therefore, a key test of different dark matter models in a way that is independent of the baryonic content of matter structures on subgalactic scales. In this chapter, we describe how galaxy-scale strong gravitational lensing observations are sensitive to the physical nature of dark matter. We provide a historical perspective of the field, and review its current status. We discuss the challenges and advances in terms of data, treatment of systematic errors and theoretical predictions, that will enable one to deliver a stringent and robust test of different dark matter models in the near future.
With the advent of the next generation of sky surveys, the number of known strong gravitational lens systems is expected to increase by several orders of magnitude. Coupled with high-resolution follow-up observations, these data will provide a key opportunity to constrain the properties of dark matter with strong gravitational lensing. 
  
\end{abstract}

\tableofcontents


\label{chapter6}

\section{Introduction} 
\label{sec:introduction}

In the standard cosmological model, 85 per cent of the total amount of matter in the Universe is made of non-baryonic dark matter \citep{Planck2018}. Evidence of the existence of dark matter spans a wide range of independent astronomical observations: the cosmic micro-wave background \citep{Planck2020}, the rotation curves of disk galaxies \citep[e.g.][]{Albada1985,Rubin1991}, the motion of galaxies within clusters \citep[e.g.][]{Zwicky1933}, gravitational lensing by galaxies and galaxy clusters \citep[e.g][]{TreuKoopmans2004,Mandelbaum2006, Clowe2006, Auger2009, Barnabe2011, Sonnenfeld2022,Shajib2022}, hot gas in galaxy clusters \citep[e.g.][]{Ettori2013}, weak lensing and galaxy clustering \citep[e.g.][]{Alam2017,Heymans2021,Abbott2022}. To this day, the physical nature of dark matter remains an unsolved problem. There is no dark matter particle within the standard model of particle physics. So far, several candidates, spanning 90 orders of magnitude in mass, have been proposed \citep{Bertone2018}. In the 1980s, weakly interactive massive particles (WIMPs) emerged as the favourite candidate for cold dark matter. The success of WIMPs stems from the fact that their existence is predicted by Supersymmetry theories \citep[][and references therein]{Fox2018}, according to which they are thermally produced with a self-interaction cross-section that leads to the correct amount of dark matter at the present day. However, decades of particle physics experiments have failed to produce a WIMP detection \citep{Arcadi2018}.

Other viable dark matter candidates include warm (WDM), self-interacting (SIDM) and fuzzy (FDM) dark matter. We refer the reader to Section \ref{sec:dm} for a more detailed description. An interesting aspect of these alternative models is that they predict a different distribution of dark matter on subgalactic scales. 

Strong gravitational lensing is sensitive to the distribution of matter and dark matter between the observer and the source. It is a purely gravitational probe and does not rely on the presence and distribution of baryons. It therefore provides a channel to observationally constrain the physical nature of dark matter. Other probes include the Lyman-$\alpha$ forest \citep[e.g.][]{Villasenor2022}, the satellite galaxies of the Milky-Way \citep[e.g.][]{Nadler2021a} and stellar streams in the Local Group \citep[e.g.][]{Erkal2017}.

This chapter focuses on galaxy-scale strong gravitational lensing as a probe of dark matter and is organized as follows. In Section \ref{sec:dm}, we give a description of the dark matter models that have so far been tested with strong gravitational lensing observations. 
We expect this list to increase in the future as theoretical predictions for structure 
formation in more dark matter models become available. In Section \ref{sec:sgl_dm}, we discuss how lensing observables (i.e. image positions, fluxes, and time delays) are affected by the distribution of dark matter on small scales. In Section \ref{sec:modelling}, we describe the process of lens modelling and how one can constrain the properties of dark matter from strong gravitational lensing observations as an inference problem. Another important aspect is the role played by degeneracies, systematic errors and unknowns, all of which are discussed in Section \ref{sec:dsu}. The chapter then continues with an historical perspective of the field (Section \ref{sec:history}), its current status (Section \ref{sec:current}) and how it is likely to evolve in the near future (Section \ref{sec:future}). We acknowledge that the last two sections may soon be obsolete. Therefore, we encourage the interested reader to complement this text with the latest relevant publications. Finally, we present our concluding remarks in Section \ref{sec:conclusions}.

\begin{figure*}
	\includegraphics[width=0.5\textwidth]{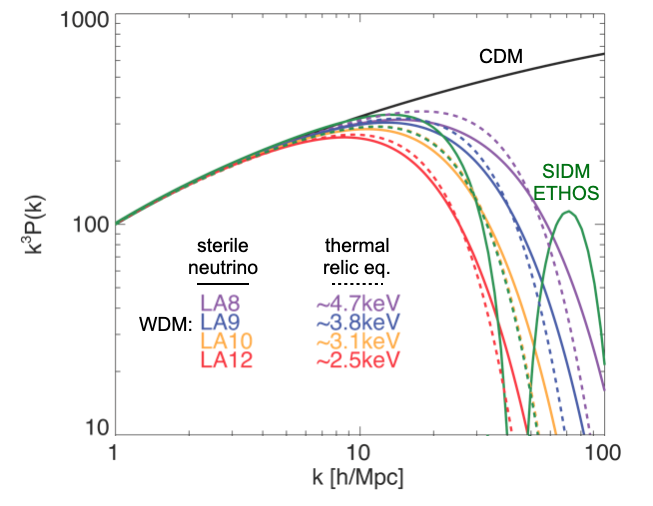}
 \includegraphics[width=0.5\textwidth]{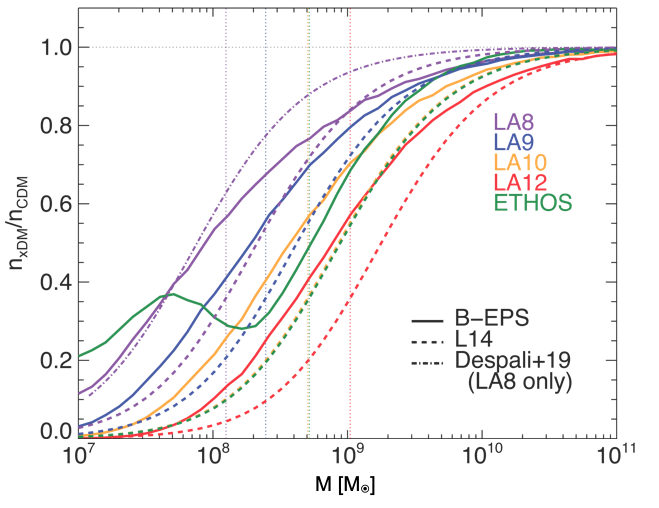}
	\caption{\label{fig:pk_dm}
     Power spectra (left) and mass functions (right) in warm and self-interacting dark matter models. In the left panel, the CDM $P(k)$ is shown in black, while different colours represent four WDM models and one SIDM model \citep[ETHOS, see][]{Vogel2016} in which the power-spectrum is modified. In the WDM case, the suppression is shown both for sterile neutrino models (solid) and their thermal-relic closest counterpart (dashed). The right panel shows instead the suppression in the halo mass function with respect to CDM, calculated with the extended Press-Schechter (EPS) formalism, or measured in simulations \citep[][]{Lovell2014,Despali2018}.  The figures are adapted from Figure 1 and 2 in \citet{Lovell2020}, respectively.}
\end{figure*}

\section{Dark matter models}
\label{sec:dm}

\subsection{Cold Dark Matter}
\label{sec:cdm}

Dark matter models where the particle is non-relativistic are described as cold. Possible candidates include WIMPs and axions \citep[e.g.][]{Feng2010}. At the same time, dark matter does not have to be a particle and primordial black holes have also been proposed as a CDM candidate \citep[e.g.][]{GreenA2021}. One of the fundamental predictions of CDM is the fact that dark matter structures form hierarchically and bottom-up: low-mass haloes form first and subsequently merge into structures of increasing mass and size. The structure and evolution of CDM haloes has been extensively studied and precisely characterised using numerical simulations \citep[see][for a review]{Zavala2019}. The collisionless nature of CDM and hierarchical structure formation result in a population of haloes with well defined properties. 

\noindent\emph{Halo mass function:} at the low-mass end, the number density $n$ of haloes is well-described, to first approximation, by a power-law distribution:
\begin{equation}
    \frac{dn}{dM}\propto M^{-\alpha}\,.
\end{equation}
For halo masses smaller than a mass of $M~\sim10^{11}M_{\odot}$ the slope is $\alpha=1.9$, and the number of structures increases with decreasing halo mass. The shape of the halo mass function is well understood from statistical arguments. These are based on the properties of the initial density field of fluctuations and the gravitational collapse process that leads to the formation of virialised haloes. This is at the basis of the formalism first introduced by \citet{Press1974}, and subsequently extended and improved by \citet{Sheth1999b}, \citet{Sheth2001b}, \citet{Tinker2008} and \citet{Despali2016}. 

\noindent \emph{Subhalo mass function:} the number density of haloes that have been accreted onto larger ones is also a power-law with a normalisation that depends on the host halo mass and redshift \citep{Giocoli2008a,Despali2017}. 
Due to the interaction between the subhaloes and the host, the former are tidally stripped and sometimes destroyed. As a consequence of this process, the total number of subhaloes in hydrodynamical simulations is reduced by a factor between 20 and 50 per cent, depending on the galaxy formation model, relative to dark-matter-only simulations \citep{Sawala2017, Despali2017}. The number density of subhaloes also changes as a function of distance from the halo centre according to an Einasto profile. However, at fixed subhalo mass, the projected number density distribution is constant with radius \citep{Xu2015, Despali2017}.

\noindent\emph{Halo mean mass density profile:} the mass density $\rho(r)$ of a dark matter halo as a function of radius $r$ is described by the Navarro, Frenk and White profile \citep[NFW,][]{Navarro1996}:
\begin{equation}
    \rho(r)=\dfrac{\rho_{s}}{\frac{r}{r_{s}}\left(1+\frac{r}{r_{s}}\right)^{2}}\,.
\end{equation}
Here, $r_s$ is the scale radius and $\rho_s$ is the density normalization. The NFW profile can also be  defined  in  terms  of  the  halo   virial  mass  $M_{\rm vir}$ \citep[i.e.  the  mass   within  the  radius, $r_{\rm vir}$,  that   encloses  a  virial overdensity $\Delta_{\rm vir}$, defined  following][]{Bryan1998}, and the virial concentration  $c_{\rm vir}=r_{\rm vir}/r_{s}$.

\noindent\emph{Subhalo mean mass density profile:} due to the tidal interactions between the subhaloes and the host halo, the mass density profiles of the former tend to deviate from a standard NFW profile and are significantly more concentrated than isolated haloes of the same mass \citep[e.g.][]{Moline2017}. For this reason, their properties are better described in terms of the peak circular velocity $V_{\rm max}$ and the corresponding $R_{\rm max}$ radius. 

In Section \ref{sec:th_un}, we discuss current uncertainties relative to the subhalo mass function and mass density profile, and how these affect the robustness of dark matter constraints from strong gravitational lensing observations.

Alternative dark matter models can affect both the number and the structural properties of haloes and subhaloes and, as a consequence, the lensing signal that they produce. In the rest of this section, we describe the most studied ones and how they differs from CDM.

\subsection{Warm Dark Matter}
\label{sec:wdm}

Dark matter particles with (close to) relativistic free-streaming velocities in the early Universe, for example, light neutrinos, are commonly described as Hot Dark Matter candidates \citep{Doroshkevich1981}. A universe predominantly made of HDM has already been ruled out by observations of the clustering of galaxies on large scales \citep{White1983}. Between CDM and HDM lies a class of dark matter models known as warm dark matter (WDM), whose candidates include the gravitino and sterile neutrinos \citep[e.g.][]{Boyarsky2009, Boyarsky2019}. They are non-relativistic, but have a non-negligibile free-streaming velocity at early times. This property leads to the suppression of the power-spectrum of the mass-density fluctuations on scales smaller than the half-mode scale $\lambda_{\text{hm}}$ (see Fig. \ref{fig:pk_dm}). The corresponding suppression in the number density of low-mass haloes relative to CDM can be expressed as follows \citep{Schneider2012,Lovell2020}:
\begin{equation}
    \frac{n_{\rm WDM}}{n_{\rm CDM}} = \left(1+ \gamma\frac{M_{\rm hm}}{M} \right)^{\beta}\,.
\label{massf_eq_2}
\end{equation}
$M_{\text{hm}}$ is the \emph{half-mode mass} and is defined as the mass-scale at which the square root of the WDM linear matter power-spectrum is 50 per cent smaller than in CDM. In practice, the half-mode mass is inversely proportional to the dark matter particle mass $m_{DM}$: the lighter the particle candidate, the stronger the suppression in the mass function at small scales. 

As a result of the increased particle velocity in WDM, structure formation is delayed to a time when the Universe is less dense. As a consequence, haloes of a given mass are not only less numerous, but also less concentrated \citep[e.g.][]{Ludlow2016}. By suppressing both halo abundance and halo concentration, the latter quantity determining the lensing efficiency of haloes, WDM models predict less small-scale perturbations to the strong gravitationally lensed images than one would expect from CDM.

\begin{figure*}
\begin{center}
\includegraphics[width=0.8\textwidth]{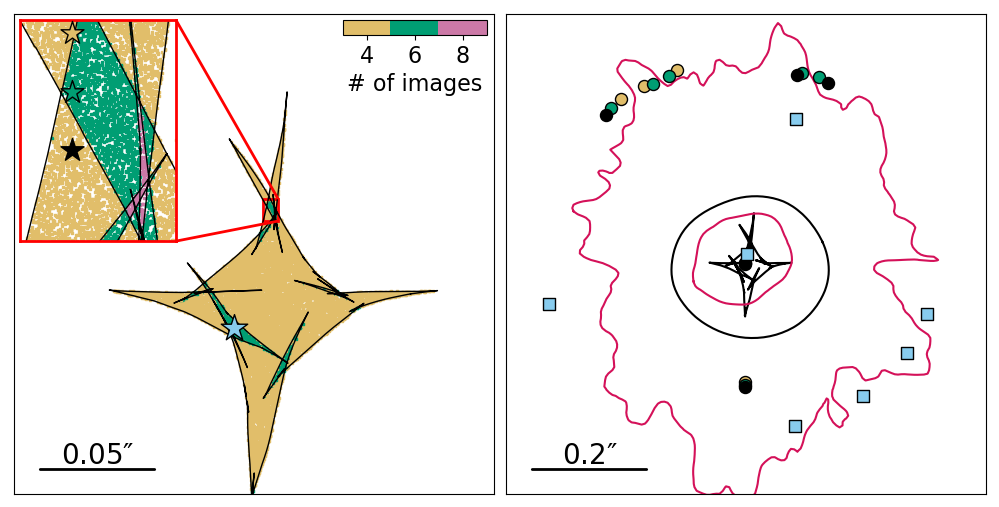}
\end{center}
	\caption{\label{fig:fdm}
    Lensed image multiplicity in FDM. The left panel shows the caustic structure of a galaxy-scale lens in FDM. The colour indicates the number of lensed images for a point source located at different positions on the source plane. The right panel shows the corresponding lensed image position on the observed plane.
    Rare hexad and octad images become more likely in FDM models. The figure is taken from \citet{Chan2020}.}
\end{figure*}

\subsection{Self-interacting Dark Matter}
\label{sec:sidm}

Self-interacting dark matter (SIDM) postulates that dark matter particles are not collisionless, but have non-gravitational interactions in which they exchange energy and momentum. The term SIDM refers to a variety of models that include elastic or inelastic scattering and a constant or velocity dependent interaction cross-section \citep{Vogel2012,Rocha2013,Kaplinghat2014,Vogel2016,Sameie2018, Robertson2018,Lovell2019,Kaplinghat2019,Vogel2019,Despali2019,Sameie2020,Kaplinghat2020,Robertson2021,Correa2021}. 

The distinctive signature of SIDM is a modification of the central mass density profile of haloes and subhaloes. The interactions between dark matter particles in high-density regions lead to a transfer of heat and the formation of a density core with a depth and size that is related to the strength of the self-interaction cross-section $\sigma$. In some cases, however, the halo and subhalo subsequently undergo a runaway contraction, also known as gravo-thermal catastrophe or core collapse. This phenomenon results in the formation of a cuspy density profile. It is accelerated by the presence of baryons \citep[e.g.][]{Feng2021} and, in the case of subhaloes, by tidal stripping \citep{Nishikawa2020}. From a strong gravitational lensing perspective, these diverse changes to the central mass density distribution imply that certain (sub-)haloes will be more efficient lenses than others \citep{Despali2019,Robertson2019,Gilman2021,Gilman2022b}. 

In addition to altering the internal structure of haloes, SIDM can also suppress the abundance of subhaloes relative to CDM through ram-pressure stripping of the subhaloes while they are accreted by the host galaxy \citep[see \ref{fig:pk_dm}) and][]{Nadler2020b,Zeng2022}. As ram-pressure stripping occurs due to self-interactions between dark matter particles bound to the subhaloes and those bound to the host, the efficiency of this mechanism depends on the amplitude of the self-interaction cross-section. For example, in velocity-independent SIDM models with elastic scattering, a very high cross-section ($\sigma\sim10$) is required. These models are currently ruled out by observations of galaxy clusters \citep{Sagunski2021,Andrade2022}.

Strong gravitational lensing, being sensitive to the amount and concentration of low-mass haloes (i.e. with a central velocity dispersion from less than $\sim$ 10 km$~\rm{s}^{-1}$ to  $\sim$ 50 km$~\rm{s}^{-1}$), provides, therefore, an independent avenue to constrain SIDM models. It allows one to constrain the self-interaction cross-section at low velocities and complements the constraints derived from galaxies and galaxy clusters \citep[e.g.][]{Loudas2022}.

\subsection{Fuzzy Dark Matter}
\label{sec:fdm}

Fuzzy dark matter is a particular form of dark matter made of \emph{ultra-light} bosons, i.e., $m_{\rm DM}c^{2}\sim10^{-22}$ eV. This particle mass is orders of magnitude smaller than that of WIMPs and WDM models. As a result, the de Broglie wavelength is larger than the inter-particle separation and waves better describe the behaviour of the FDM field. This effect leads to a series of distinctive phenomenologies with respect to the other dark matter models so far considered \citep[see][for a review]{Hui2021}. For example, numerical simulations, which model the full non-linear evolution \citep{Schive2014a,May2021}, show that the wave-like behaviour (e.g. interference effects) of FDM leads to the formation of a soliton core at the centre of haloes and density granules on scales smaller than a kpc. 

Similarly to WDM, there is a cut-off in the FDM transfer function at small scales, though via a different mechanism that is dependent on the de Broglie wavelength rather than a free-streaming length\footnote{Note, however, that the wave-like nature of ultra-light dark matter particles results in the FDM power-spectrum to briefly exceeds that in CDM on $\mathcal{O}(\mathrm{kpc})$ scales that correspond to the mean de Broglie wavelength.}. As a consequence, from a strong gravitational lensing perspective, FDM models are expected to lead to fewer perturbations of the lensed images compared to CDM. However, the granular structure in the halo density profile of lens galaxies has been shown to lead to a distinct new source of small-scale perturbations of the lensed images, which is unique to FDM models \citep[see Fig. \ref{fig:fdm} and][]{Chan2020,Laroche2022,Powell2023,Amruth2023}. 

\section{Strong lensing as a probe of dark matter}
\label{sec:sgl_dm}

Changes to the matter distribution on subgalactic scales within lens galaxies and along their line of sight leave a gravitational imprint on the strong lensing observable. Here, we briefly describe the nature and strength of this effect. In particular, we discuss the changes that low-mass haloes induce on the lensing potential, its first and second derivative, and how they affect the observed time-delay, image positions and magnifications, respectively. For an historical perspective of the field, we refer the reader to Section \ref{sec:history}.

\begin{figure*}
\begin{center}
    	\includegraphics[width=0.8\textwidth]{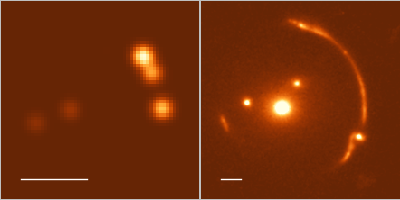}
\end{center}
	\caption{\label{fig:lens_montage}
	Two examples of galaxy-scale lenses with a compact (left) and extended (right) background source. On the left panel, the gravitational lens system B2045+265 shows a strong flux-ratio anomaly which is inconsistent with a smooth lens mass distribution \citep{Fassnacht1999}. On the right, the gravitational lens system SDSSJ120602.09+514229.5, also known as the Clone, displays a distorted arc due to the presence of a small luminous satellite galaxy \citep{Vegetti2010a}.}
\end{figure*}

\noindent\emph{Image magnification:} The largest effect produced by local fluctuations in the lensing mass density distribution (either in the form of subhaloes, field haloes or FDM granules) is a change in the relative magnification of unresolved lensed images (see Fig. \ref{fig:lens_montage} for an example), and the creation of so called flux-ratio anomalies. These anomalies could also be related to micro-lensing by stars (see Chapter 5) or propagation effects such as free-free absorption at long wavelength \citep{Mittal2007} and dust extinction \citep{Eliasdottir2006} in the lens galaxy. The observed ratios are also affected by intrinsic and extrinsic variability of the background source \citep{Koopmans2003,Biggs2023}. As such, they can only be used as a probe of dark matter with observations at wavelengths where the angular size of the lensed object is larger than the scale of the micro-lenses, and at which dust and free electron absorption is low. Since flux-ratio anomalies are related to a local change of the second derivative of the lensing potential, they are an effective tool to detect some of the lowest mass perturbations. 
Lenses with small opening angles in the fold and cusp configurations (i.e. where the source lies on the cusp or fold of the caustic curves) are the most sensitive to the perturbations of the lensing potential and least sensitive to intrinsic variability. In these cases, the following $R_{\rm fold}$ and $R_{\rm cusp}$ relations can be used to quantify the strengths of the anomaly:  
\begin{equation}
    R_{\rm fold} = \frac{\mu_A + \mu_B}{|\mu_A| + |\mu_B|}\,,
 \label{equ:fold}   
\end{equation}
and
\begin{equation}
     R_{\rm cusp} = \frac{\mu_B + \mu_A + \mu_c}{|\mu_A|+|\mu_B|+|\mu_C|}\,.
      \label{equ:cusp}   
\end{equation}
Here, $\mu_{A,B}$ in $R_{\rm fold}$ and $\mu_{A,B,C}$ in $R_{\rm cusp}$ are the magnifications of the merging double and triplet images, respectively. As the opening angles between the images A and C ($\Delta \phi$) in the cusp configuration and A and B ($\phi_1$) in the fold configuration approach zero, so do $R_{\rm fold}$ and $R_{\rm cusp}$, when the lens mass distribution is smooth and the background source is a point-source.
However, it is worth noting that astrophysical sources of emission, whether they be the accretion disk, the narrow emission line region or the warm ($>$ 50 K) dust torus around a black hole or the relativistic jet that they produce are not point-like, but have some angular scale.

\noindent\emph{Image positions:} The second largest effect is a local change of the relative positions of the multiple lensed images and the creation of so called astrometric perturbations. These anomalies are related to a local change of the first derivative of the lensing potential. Therefore, they can only be produced by a gravitational perturbation and cannot be caused by micro-lensing nor dust. In the case of extended sources, astrometric anomalies appear as perturbations to the surface brightness distribution of highly magnified arcs and Einstein rings, and are sometimes referred to as surface-brightness anomalies (see Fig. \ref{fig:lens_montage} for an example). 
The level of perturbations that can be detected using a surface-brightness anomaly is set by the quality of the data and structure of the source surface brightness distribution \citep[see Section \ref{sec:current} and][]{Despali2022}. For example, a subhalo with a mass of 10$^6 M_\odot$ will produce distortions on angular scales of a couple of milli-arcsecond.
 
\noindent\emph{Image time delay:} The weakest effect is a change of the lensing potential itself. In the case of multiply imaged quasars with a flux-varying source, this phenomenon is observed as a perturbation to the time delay (see Chapter 7) between the multiple images \citep{Keeton2009,Cyr-Racine2016,Gilman2020c}. Unlike flux-ratio anomalies, time-delay anomalies are not affected by dust extinction in the lens galaxy. However, they suffer from stellar micro-lensing which induces time-delay changes of the order of days \citep{Tie2018}. Time-delay anomalies due to subhaloes are typically of the order of a fraction of a day. At present, this is smaller than the typical time-delay uncertainty \citep[equal or larger than a day, e.g.,][]{Fassnacht2002}. To be an effective probe of the nature of dark matter, time-delay anomalies require, therefore, sensitive measurements with a high observing cadence.

\noindent\emph{Image polarisation angle:} The presence of a field of Axion-Like Particles (ALPs) and its interaction with photons leads to the so-called phenomenon of birefringence whereby the polarisation angle of a linearly polarised source gets rotated. If the field oscillates over time, so does the change in the polarisation angle in a way that is related to the mass of the particles. Due to gravitational time delay, the multiple images of strongly lensed sources experience a different level of rotation, leading to differential birefringence, which can be used to probe the particle mass and it coupling with photons \citep{Basu2021}.

\section{Lens modelling} 
\label{sec:modelling}

The process of constraining the properties of dark matter with strong gravitational lensing is best understood as a (hierarchical) Bayesian inference problem with the following unknowns: the intrinsic properties of the source $\vec{s}$, the parameters of the main lens(es) mass distribution $\vec{\eta}$, and the amount and properties of low-mass haloes or fuzzy dark matter granules $\vec{\eta}_{\rm pert}$. These have to be simultaneously inferred from the observed data $\vec{d}$ to which they are related in a statistical sense via the following posterior distribution:
\begin{equation}
P(\vec{s},\vec{\eta},\vec{\eta}_{\rm pert}|\vec{d}) = \frac{P(\vec{d}|\vec{s},\vec{\eta},\vec{\eta}_{\rm pert}) P(\vec{s},\vec{\eta},\vec{\eta}_{\rm pert})}{P(\vec{d}|M)}\,.
\label{eq:posterior}
\end{equation}
In the numerator, the first term is the Likelihood function and the second is the prior on the parameters of the model $M$. The term in the denominator is the marginalised Likelihood (also known as Bayesian evidence). A detailed description of each ingredient is provided in the following sections. The equation above is an exact expression of the inference problem at hand, and it encodes how the different components are related to each other. However, depending on the modelling approach (see Section \ref{sec:modelling_approaches}), one may not make direct use of this posterior probability and instead adopt simplifying assumptions to make the inference problem more tractable.

\subsection{The Likelihood}
\label{sec:likelihood}

The Likelihood function returns the probability of observing the data, given a choice of model. In the context of strongly lensed unresolved sources, the data $\vec{d}$ is the set of image fluxes and positions. For extended sources, the data is the surface brightness distribution in each image pixel in optical observations and a set of visibilities for interferometric ones. In the case of radio polarised emission, the data is given by the visibilities of the coherency vector. In studies that combine multiple data-sets of the same lens system at different wavelengths, the data can be thought of as a concatenation of the different observations, which originate from the same lensing potential, but with a different surface brightness distribution of the source. In many cases, the Likelihood function is well approximated by a Gaussian distribution with uncorrelated noise\footnote{However, optical images may have correlated noise as a result of the pixel re-sampling using, for example, the drizzle image processing.}.

\subsection{The source}
\label{sec:source}

Sources in gravitational lens systems can be of a large variety. In the following, we simply classify them into two categories: compact and extended sources. AGN and QSOs are examples of compact sources. Extended sources include galaxies and radio jets. 

\noindent\emph{Compact sources:} Traditionally, for gravitationally lensed sources that appear unresolved, either a point or a Gaussian source is assumed. Free parameters of the model are therefore the source position and flux, and in the latter case also the source size. Both assumptions can introduce systematic errors in the dark matter inference that are discussed in more detail below.

\noindent\emph{Extended sources:} Typically, one of the following methods is used to describe the surface brightness distribution on the source plane: an analytical (e.g. a S\'{e}rsic) profile, a pixellated model, a basis function regression (e.g. starlets and shapelets) or a deep generative model. Here, we briefly discuss the main strengths and weaknesses of each approach.

The advantage of analytical models is that the number of free parameters is small, and the modelling procedure is fast. The main drawback is that such simplistic sources are an unlikely representation of lensed star-forming galaxies, which are often clumpy and irregular \citep[e.g.][]{Ritondale2019a}. 

Pixellated sources have enough freedom to fit complicated light distributions well. However, as the lensing problem is poorely constrained, they require a regularising prior \citep[e.g.][]{Suyu2006,Birrer2015,Vernardos2022,Galan2022}. The choice of prior is non-trivial and may not necessarily be physically motivated. A possible solution is the introduction of hyper-priors \citep{Rizzo2018,Vernardos2022}, which have the advantage of retaining the freedom of a pixellated source while imposing physically-meaningful constraints. Another limitation of free-form models is that they are challenging from a computational perspective, especially for high-resolution interferometric data \citep{Hezaveh2016b, Powell2021}. One more aspect to consider is that these models require constructing a regular or adaptive grid on the source plane. We refer the reader to \citet{Tagore2014} and \citet{Nightingale2015} for a detailed discussion of biases related to different choices of source discretisation and regularisation schemes.

Recently, machine learning techniques have been introduced as a model for source galaxies \citep[e.g.][]{Chianese2020, Adam2022, Karchev2022}. These approaches can overcome some of the above limitations. However, their performance is sensitive to the choice of training data. It is currently unclear how to generate large training samples of realistic lensed galaxies at high angular resolution. It is, for example, unlikely that nearby galaxies are a good description of the high-redshift population.
Similarly, lensed and unlensed galaxies at the same redshift are not observed on the same angular scales. \citet{Holzschuh2022}, have shown how generative models can be used to create arbitrarily large samples of source galaxies from hydrodynamical simulations. However, it is unclear how well these simulated galaxies represent the population of lensed sources. As for pixellated sources, deep learning techniques also suffer from biases related to the data discretisation.

Independently of the chosen model, a certain level of degeneracy exists between structures in the source and structures in the lensing potential. Therefore, any assumption on the source light has important consequences for the inference of dark matter with strong gravitational lensing. We provide a more detailed discussion of this issue in Section \ref{sec:dsu}.

\subsection{The lensing potential}

In this chapter, we focus our attention on galaxy-scale lenses. While galaxy clusters could also provide constraints on small-scale fluctuations in the dark matter distribution, most studies so far have focused on smaller scale deflectors. This choice is related to the challenge of modelling the complicated mass distribution of galaxy clusters with enough precision.

\subsubsection{Main lens}
\label{sec:macro_model}

For galaxy scale lenses, the main deflector is typically a single galaxy, most commonly but not limited to a massive early-type. For many years, the most common parametrization for its mass density distribution was a single power law (SPL) or a singular isothermal (SIE) elliptical profile, plus external shear. This choice was motivated by the analysis of relatively large samples of lens galaxies \citep[ e.g.][]{Koopmans2009,Barnabe2011}. However, it is worth noting that no existing sample of strong lenses is representative of the actual distribution of observable lenses in the Universe. 

An important question is the level of complexity in the mass distribution of lens galaxies and how it affects the constraints on dark matter. For example, deep Near-Infrared (NIR) observations have revealed the presence of a disk component in two lenses that were previously assumed to be purely elliptical. It was also shown that these disks have a non-negligible lensing effect and account for most of the observed flux ratio anomaly \citep{Hsueh2016,Hsueh2017}. Similarly, \citet{Spingola2018} and \citet{Powell2022} have shown with VLBI observations that an SPL is a good description of the lens mass distribution of MG J0751+2716 only down to scales of a few milli-arcseconds. Below these scales more complex angular and radial structures become important. In light of these and other results (see Section \ref{sec:dsu}), more recent analyses have included the effect of multipole mass moments, such as boxiness and diskyness, as well as that of nearby satellite galaxies. 
The degeneracy between different forms of complexity in the lensing potential (e.g. subhaloes versus disks or multipoles) is a source of systematic error in the inference of dark matter. We discuss this problem and possible solutions in Section \ref{sec:dsu}.

\subsubsection{Low-mass haloes}
\label{sec:low_mass_haloes}

In the following, we refer to subhaloes and field haloes collectively as low-mass haloes. The parameters describing this population are: the number of objects, their masses, positions, redshifts and mass density profiles. Different assumptions regarding these quantities can be found in the literature. Here, we provide a description of those that are most frequently used.

\noindent\emph{Analytical:} most commonly, low-mass haloes are modelled as spherical systems with a Pseudo-Jaffe \citep[PJ, e.g.][]{Dalal2002} or a (truncated) NFW \citep[e.g.][]{Gilman2019, Hsueh2020} profile. The mass-concentration relation of the NFW depends on the dark matter properties and is either taken from numerical simulations \citep[e.g.][]{Gilman2019} or is a free parameter of the model \citep[e.g.][]{Gilman2020b}. In fully forward models, the number of objects as a function of their redshift and mass are drawn from a Poisson distribution with an expectation value derived from the halo and subhalo mass functions. The latter are set by the dark matter model and are taken from numerical simulations (see Section \ref{sec:dm}). On each redshift plane, the projected positions of low-mass haloes are generally assumed to be uniformly distributed, as motivated by numerical simulations \citep{Xu2015}.

\noindent\emph{Pixellated:} low-mass haloes are described as linear local corrections to the lensing potential \citep{Koopmans2005,Vegetti2009a}. Individually detected objects are identified as positive convergence corrections 
of the otherwise smooth lens mass distribution. As for pixellated sources, a regularising prior for the potential corrections must be defined \citep[e.g.][]{Vernardos2022}. Due to its free-form nature, this approach does not make a priori assumptions on the number and mass density profile of the low-mass haloes. Typically, the connection to specific dark matter models is done a posteriori as described in Section \ref{sec:modelling_approaches}. 

\noindent\emph{Gaussian Random Field:} low-mass haloes are represented by a Gaussian Random Field \citep[GRF, e.g.][]{Hezaveh2016a,DiazRivero2018a, DiazRivero2018b,Chatterjee2018,Cyr-Racine2019,Bayer2023a,Bayer2023b}. The corresponding power-spectrum, which is by construction well represented by a power-law, then carries information on the low-mass haloes abundance, mass function and density profile. Hence, these quantities do not have to be assumed a priori. As the assumption of Gaussianity only holds for the very low mass haloes that appear in great number, the more rare and massive objects have to be individually detected and separately treated. The connection to the properties of dark matter is done a posteriori and in terms of the power-spectrum (see Section \ref{sec:modelling_approaches}).

\subsubsection{Fuzzy Dark matter granules and subhaloes}

In FDM cosmologies, the small-scale structure of galaxy-scale haloes and their subhalo populations are markedly different from their CDM or WDM analogues. The main difference is that the main dark matter halo exhibits wave intereference on $\sim$kpc scales due to the ultra-low mass of the dark matter particle; this gives rise to $\mathcal{O}(1)$ fluctuations in the halo density, which are commonly termed ``granules''. FDM haloes that are self-consistent with regard to the governing Schr\"{o}dinger-Poisson equations can be obtained via numerical simulation \citep{Schive2014b,May2022} or direct construction of wave-function eigen-modes \citep{Yavetz2022}. However, for the practical purpose of gravitational lens modeling, a faster analytic prescription is often preferred. To this end, \citet{Chan2020} and \citet{Kawai2022} derive statistical properties of FDM granules that can be used to quickly generate perturbations to a smooth lensing potential, which are consistent with an FDM halo. \citet{Laroche2022} implement this approach by randomly placing a large population of Gaussian density profiles in the lens, while \citet{Powell2023} use a Fourier-space approach to achieve a similar result.

\subsection{Modelling approaches}
\label{sec:modelling_approaches}

\subsubsection{Semi-linear}

\citet{Warren2003} presented a lens modelling approach for the analysis of data with an extended source in which the latter is pixellated (see Section \ref{sec:source}). The most probable a posteriori (MAP) source and main lens parameters are inferred from the posterior probability in Eq. (\ref{eq:posterior}). In the case of quadratic prior distributions, the MAP source is obtained by solving a linear system \citep[hence the term semi-linear - see][for examples of a semi-linear techinique with a basis function regression model for the source]{Birrer2017,Galan2022}. The Bayesian evidence (denominator in  Eq. \ref{eq:posterior}) is used to compare different models for the lensing potential and different choices of regularizations.

\citet{Koopmans2005} introduced the so-called gravitational imaging technique for the detection of low-mass haloes with galaxy-galaxy lensing. In this approach, the source is pixellated and the low-mass haloes are described as linear, pixellated corrections to the analytical main lensing potential (see Section \ref{sec:low_mass_haloes}). The methodology was fully embedded in the framework of Bayesian statistics with an adaptive source by \citet{Vegetti2009a} and extended to the 3D (one frequency and two spatial dimensions) and interferometric domain respectively by \citet{Rizzo2018} and \citet{Powell2021}. Recently, \citet{Vernardos2022} have further extended the original method by \citet{Koopmans2005} to include physically motivated priors. Due to the pixellated nature of the potential corrections, the gravitational imaging methodology does not require any assumption on the number, mass and position of low-mass haloes. Indeed, \citet{Vegetti2009a} have explicitly shown that more than one subhalo can be identified (provided that they have an effect on the lensed images), and \citet{Dhanasingham2022} have introduced a formalism to differentiate between subhaloes and field haloes based on the two-point function of the effective deflection angle field. Moreover, the potential corrections are not limited to capturing the effect of low-mass haloes. As shown by \citet{Barnabe2009}, \citet{Ritondale2019b} and \citet{Galan2022} they can be used to identify components in the lensing potential that are not captured by the main parametric lens model. Indeed, the freedom allowed to the lensing potential is one of the main advantages of this approach as one can directly identify and differentiate different forms of complexity. One disadvantage is that the method is not fully forward, and two more steps are required to derive constraints on the properties of the dark matter: assessing the statistical relevance of detections and non-detections, and the interpretation of these in the context of theoretical predictions. 

\noindent\emph{Detections:} \citet{Vegetti2010b}, \citet{Vegetti2012} and \citet{Ritondale2019b} have introduced the following criteria to define the detection of a low-mass halo as statistically robust. (i) The mass and position of the pixellated convergence corrections have to be consistent with those inferred from an analytical description of the low-mass halo. The latter is inferred from the posterior distribution with a non-informative prior on the object mass and position, and a given choice of mass density profile. (ii) The model that includes low-mass haloes is preferred over the smooth one with a Bayes factor of at least 50. 
Under the assumption of statistical Gaussian errors, this difference in Bayesian evidence corresponds roughly to a $\geq$10-$\sigma$ detection. While this may sound overly conservative, this choice is made due to the presence of unaccounted-for systematic errors that can result in false positives at lower significance levels. For example, \citet{Ritondale2019b} have found that the rate of false positives changes from 40 per cent with a 4-$\sigma$ level threshold to 20 per cent at the 5-$\sigma$ level. At the same time, the rate of false negatives decreases from 60 per cent with a 4-$\sigma$ level threshold to 30 per cent at 6$\sigma$. Recently, this issue was more systematically quantified by \citet{Nightingale2022}. We discuss their work in more detail in Section \ref{sec:dsu}. It should be noted that false detections are not limited to this lens modelling approach. They are intrinsic to the general problem of detecting low-mass haloes with lensing and their degeneracy with other aspects of the lens inference problem. If anything, requiring that the free-form and the analytical model are consistent with each other can significantly mitigate some of these issues and reduce the incidence of false positives.

\noindent\emph{Non detections:} \cite{Vegetti2014} and \citet{Ritondale2019b} quantify the statistical relevance of the non-detections using the so called sensitivity function. For each pixel on the image plane the sensitivity function returns the lowest subhalo mass that could have been detected with a Bayes factor of 50. One of the main draw-backs of the sensitivity function is that, depending on the size of the data, it can be computationally expensive to evaluate. For each pixel on the image plane one has to calculate the Bayesian evidence of the model with a subhalo of a given mass. Hence marginalizing over redshift, number, mass and mass density profile could be practically unfeasible. \citet{ORiordan2023} have recently shown that in principle one can successfully overcome these computational limitations by calculating the sensitivity function with machine-learning approaches. In Section \ref{sec:current} we will discuss in more detail what sets the sensitivity of a given data set to the presence of low-mass haloes of a given mass.

\noindent\emph{Dark matter constraints:} From the sensitivity function, one can then interpret detections and non-detections within a given dark matter model. This is obtained by calculating the posterior probability of the dark matter particle mass assuming that the number of low-mass haloes has a Poisson distribution with expectation value given by the halo and subhalo mass function from that specific dark matter model \citep{Vegetti2014,Vegetti2018,Ritondale2019b,Enzi2021}.\\

\citet{Hezaveh2016a} have introduced a new formalism in which the presence of subhaloes in the lens galaxy is modelled as a GRF (see Section \ref{sec:low_mass_haloes}). From the lensing observations, one constrains the power spectrum of projected density fluctuations, which amplitude and shape can then be a-posteriori compared to predictions from different dark matter models. The advantage of this approach is that no a priori assumption is made on the properties of the low-mass haloes as these can be inferred from the analysis itself. Indeed \citet{DiazRivero2018a} have shown that the amplitude and shape of the power-spectrum are sensitive to the abundance, mass density profile, and concentration of subhaloes. The main disadvantage is that the assumption of a GRF only holds for the lowest-mass objects $M < 5 \times 10^7 M_\odot$ \citep{Hezaveh2016a}. Hence, the larger ones have to be first individually identified and then explicitly included in the mass model. Moreover, in its current implementation, this approach does not allow one to identify other forms of complexity in the lens mass distribution (these are unlikely to be well described by a GRF), which then introduce a systematic bias on the dark matter inference \citep[e.g.][]{Bayer2023b}. 

\subsubsection{ABC}
\label{sec:abc}

Approximate Bayesian Computation \citep[ABC,][]{Rubin1984,Diggle1984, Tavare1997, Turner2012, Liepe2014} is an algorithm rooted in Bayesian statistics to circumvent the direct calculation of intractable Likelihood functions. As such, it enables inference analysis based on simulated data sets computed in a forward fashion.
ABC methods follow a common general process: (i) emulate the data many times with different underlying target parameters as well as noise realizations; (ii) compress the difference between the simulation and the data in a set of summary statistics to provide a metric distance between the simulation and the data; (iii) accept the proposed simulations and their underlying parameters if the distance metric is within a certain threshold $\epsilon$ and (iv) the accepted samples can be interpreted as posterior distributions with the prior being the draws of the simulator.
If the acceptance criteria converges to the identical matching of the simulation and data, $\epsilon \rightarrow 0$, the accepted sample is identical to the posterior from the exact Likelihood expression applied on the summary statistic. 

An accurate ABC inference is made of three main components. The first one is accurate simulations including all relevant aspects affecting the data (or more specifically the summary statistic). The second ingredient is a summary statistic that captures significant aspects of the signal of interest. Finally, one needs a sufficient number of simulations such that a narrow acceptance criteria (small $\epsilon$) can be chosen that leads to convergence and an accurate posterior prediction. One of the main advantages of an ABC approach is that, in contrast to some machine-learning-based methods (see Section \ref{sec:machine_learning} for more details), the summary statistic is explicit and is a means of understanding the impact of possible sources of systematic errors. However, the choice of summary statistic is arbitrary, potentially leading to a significant loss of information and constraining power. Specifically, the choice of summary statistic and level of data compression determine the number of simulations needed for the ABC process to converge, potentially resulting in a loss of computational efficiency relative to machine learning approaches.

In the context of inferring the properties of dark matter with strong gravitational lensing, \cite{Birrer2017} were the first to propose an ABC method. In particular, they made use of an elaborate summary statistics consisting of a substructure scanning approach and the map of relative Likelihood (Section \ref{sec:likelihood}) values in the reconstruction with and without a substructure. This scanning mechanism was meant to filter signal and not absorb other spurious effects between the simulations and the data. The methodology was designed for strong gravitational lens systems with an extended source, which was modelled with a basis function regression method in the form of shapelets. We refer the reader to \citet{Enzi2020}, \citet{He2022}, \citet{Bayer2023a} and \citet{Bayer2023b} for other examples of summary statistics commonly used when modelling strongly lensed galaxies.
\citet{Gilman2018}, \citet{Gilman2019} and \citet{Gilman2020a} have made use of an ABC approach to analyse the fluxes and positions of strongly lensed quasars. In this case, as the size of the data $\vec{d}$ was small, no compression was required and the summary statistics was set to the difference in the flux ratio of the multiple images between the simulation and the data. 

As the ABC is a forward method, subhaloes and field haloes are typically described by analytical mass profiles and have properties statistically drawn as described in Section \ref{sec:low_mass_haloes}.

\subsubsection{Trans-dimensional}
\label{sec:trnsdm}

One of the challenges in constraining the properties of dark matter with strong gravitational lensing is that the number of low-mass haloes in any given lens system is unknown. As a consequence, the model describing the low-mass halo population and its properties needs to have a variable and a-priori unknown number of free parameters. An additional challenge is that most of the low-mass haloes will be at or below the detection limit. \citet{Brewer2016} and \citet{Daylan2018} have proposed the use of trans-dimensional Bayesian inference approaches \citep{Green1995,Green2003} to overcome these problems. 

These methods apply probabilistic cataloging to images of strongly lensed systems \citep{HoggLang2010, Brewer2013, Brewer2016, Daylan2017, Portillo2017} and output an ensemble of probability-weighted (sub-)halo catalogs providing a good fit to data. The prior distribution for the properties of the low-mass haloes is specified hierarchically, so that their mass function is a natural output of the method. Sampling of the posterior can be done with reversible jump Markov Chain Monte Carlo with differences in the explicit sampling methods, such as Diffusive Nested Sampling \citep[DSN,][]{Brewer2016} or PCAT \citep{Daylan2018}. One can also evaluate the marginal Likelihood of the model, including over the unknown number of low-mass haloes, and the source and lens properties. In most applications, both the source and the lens are described by analytical models. However, there is in principle no limitation to couple a trans-dimensional treatment of low-mass haloes with, for example, pixellated sources. 

\subsubsection{Machine Learning}
\label{sec:machine_learning}

In recent years, it has been shown that neural networks can efficiently estimate the parameters of strong lens models directly from observations \citep{Hezaveh2017,Levasseur2017,Schuldt2021}. Motivated by these results, a number of works have explored machine learning methods for identifying subhaloes in mock observations \citep{Ostdiek2020, Lin2020, DiazRivero2020}. Others have shown that subhalo summary statistics can be directly estimated from mock observations \citep{Brehmer2019,Alexander2020a,Alexander2020b,Varma2020,Alexander2021,Vattis2021,Ostdiek2022}. However, these works do not provide any rigorous uncertainty estimates.

To address this issue, \citet{Brehmer2019} proposed an approach relying on Neural Ratio Estimators (NREs). NREs transform the problem of inferring the value of a continuous variable into a classification problem between two sets. Using this, it is possible to calculate the posterior of a parameter as the ratio of the Likelihood, multiplied by the prior, and the evidence. This allows for better uncertainty estimates. However, the approach of \citet{Brehmer2019} is extremely data-hungry, making its scaling to realistic data non-trivial. A number of variants of NREs have been proposed to solve this problem \citep{Cranmer2020,Coogan2020,Montel2022}.

Uncertainties on the subhalo population can also be obtained similarly to those of the macro-model as proposed by \citet{Levasseur2017}. \citet{Wagner-Carena2022} showed that it is possible to infer the parameters of the subhalo mass function over a population of gravitational lenses, in a Bayesian hierarchical formalism. They trained a neural network that directly predicts a parameterized distribution approximating the posterior of the low-dimensional macro-model parameters and subhalo mass function normalization \citep[see also][]{Vernardos2020}. 
The issue of how to consistently incorporate realistic sources of biases in this inference such as complex selection functions, for example, selection from a lens-finding neural networks, has been discussed by \citet{Legin2022}.

Together, these works have shown that neural networks can extract low dimensional information about the density fields, that is, the parameters of the mass function or $m_{\mathrm{WDM}}$. The main limitation of these methods, however, is that they are only tractable for the inference of low-dimensional representations, implicitly marginalizing over all nuisance parameters that are not explicitly estimated. At first glance, implicit marginalization appears appealing because the marginal posterior of the dark matter model parameters is the desired outcome. However, it hinders reproducibility and confirmation of the results by alternative, more traditional approaches. Machine learning models do not predict, at least for now, the actual distribution of density in the foreground, or the surface brightness of the background source. One cannot therefore compare their predictions with the data by ray-tracing through the model favoured by the neural network.

If, for example, excess power is detected in the mass function, it would be difficult to identify the specific feature in the data that accounts for it. Investigating whether a neural network's preference for a given dark matter model comes from a massive subhalo or a population of low mass subhaloes would be intractable in most machine learning frameworks developed to date. The lack of explicit predictions of physical features like density or brightness distributions, makes it difficult to verify measurements of low-dimensional substructure statistics by neural networks using more traditional methods.

One of the central difficulties in modelling the physical features of strong gravitational lenses, either with traditional analysis methods or with deep learning, has been the difficulty of defining priors over such high dimensional spaces. Recent advances in denoising diffusion-based models are now making this previously-intractable problem possible \citep{Song2020}. Further, \citet{Adam2022} have shown that a neural network can learn the score of the prior over background source images, learned from unlensed high-resolution images of galaxies. By adding the Likelihood score to this learnt prior score and using a reverse-time stochastic differential equation solver, the authors can obtain samples from the posterior over background source pixels. Generalizing the framework of denoising score-based models to solve highly non-linear problems such as the reconstruction of density maps in strong gravitational lenses is an active area of research.

\section{Degeneracies and systematic errors}
\label{sec:dsu}

When assessing the robustness of strong gravitational lensing constraints on dark matter it is important to consider the effect of degeneracies and systematic errors. Some of these are related to the measurement at hand or unsolved theoretical questions and affect both compact and extended sources. Others are lens modelling dependent. All of them are discussed in this section.

\subsection{Degeneracy with complex macro-models} 
\label{sec:dsu_macro}

\begin{figure*}
\begin{center}
\includegraphics[width=0.9\textwidth]{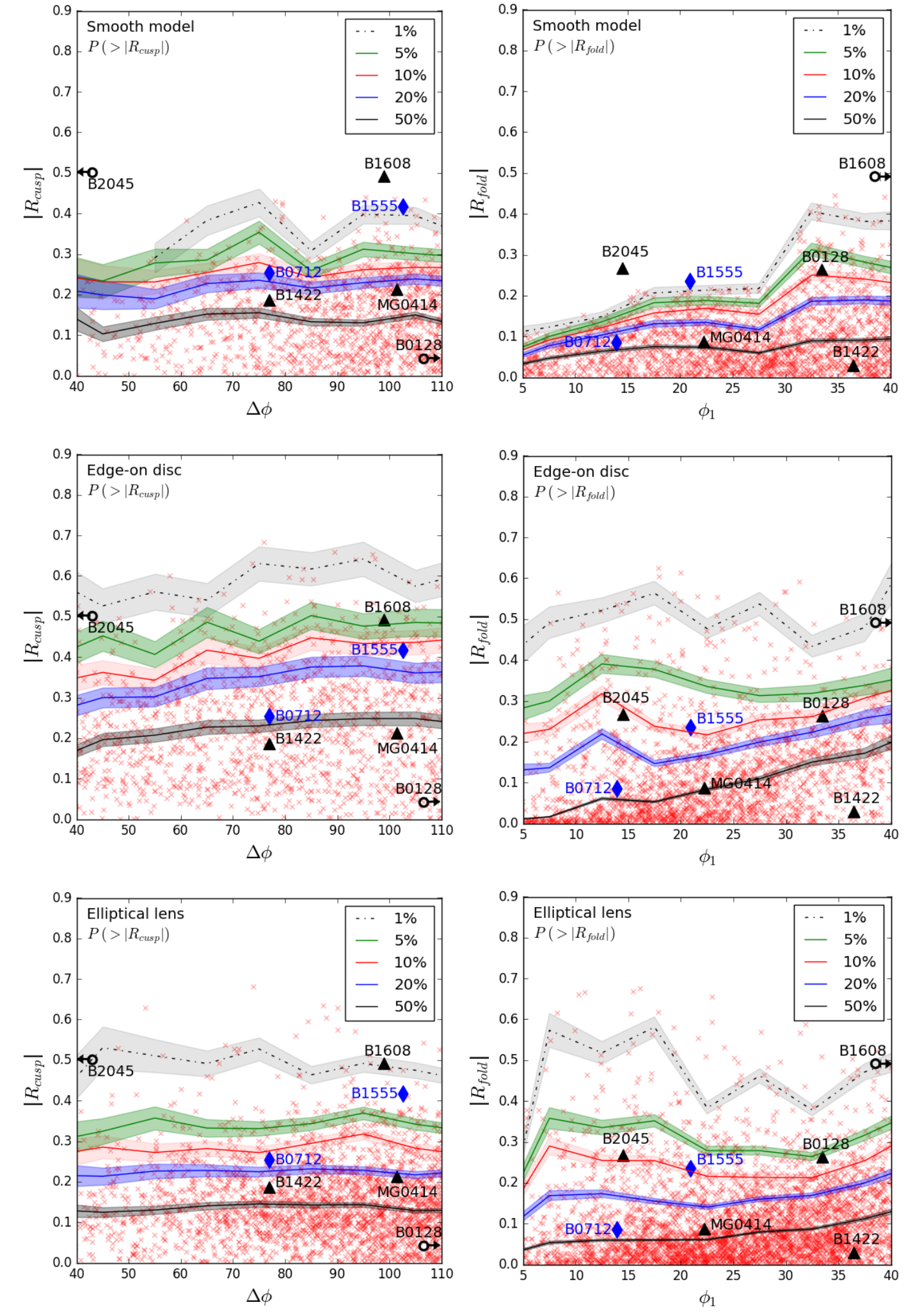}
	\caption{\label{fig:FA_baryons}
	Strength of the flux-ratio anomaly for a cusp (left) and fold (right) configuration arising from different mass distributions: an elliptical smooth power-law (top), an edge-on stellar disk (middle) and an early-type galaxy (bottom) selected from the Illustris simulation \citep{Vogel2014a}. The curves represent 1, 5, 10, 20, and 50 per cent probabilities to find values of $|R_{\rm fold}|$ and $|R_{\rm cusp}|$ larger than a given value for fixed opening angles $\phi_{\rm 1}~(\circ)$ and $\Delta\phi~(\circ)$. The panels are taken from Figures 5, 6 and 7 by \citet{Hsueh2017}.} 
 \end{center}
\end{figure*}

For any analytical model of the main deflector, a degeneracy exists between the macro-model parameters (e.g. Einstein radius and external shear) and the presence and properties of low-mass haloes. Its strength depends on the data signal-to-noise ratio (SNR) and angular resolution, and the properties of the low-mass haloes themselves.
For low-mass haloes that can be individually detected, the more concentrated their density profile and the closer to the lensed emission their position, the smaller the degeneracy with the lens properties \citep{Despali2022}. 

At the same time, real galaxies are unlikely to be simple elliptical objects and are expected to have complex radial and angular structures \citep[e.g.][]{Spingola2018, Powell2022,Vyvere2022}. Unmodelled components in the lensing potential are likely to lead to an overestimation of the amount of low-mass haloes or FDM granules, resulting in a biased inference on the properties of dark matter. 
For example, \cite{Xu2015} have shown that the level of flux-ratio anomalies in eight strongly lensed quasars from the CLASS survey \citep{class1,class2} is too large to be solely explained by low-mass haloes from CDM. They concluded that other forms of departure from a power-law macro-model may be partly contributing to the observed signal. Thanks to deep NIR observations of the CLASS gravitational lens systems B1555+375 and B0712+472, 
\citet{Hsueh2016} and \citet{Hsueh2017} were able to show that previously undetected stellar disks are responsible for most of the observed flux-ratio anomaly in these systems. Indeed, using mock simulated data, \citet{Gilman2017} and \citet{Hsueh2018} have reported that baryonic structures in the main lens can lead to an increase of the probability of large flux-ratio anomalies of between 10 and 20 per cent, depending on the lens galaxy morphology (see Fig. \ref{fig:FA_baryons}). 
Similarly, from the analysis of the the Cosmic Horseshoe lens system, \cite{Brewer2016} concluded that the inferred substructure population might have partially mimicked larger scale components not included in the primary lens model. 
\citet{He2022} have found that deviations from an elliptical shape that are not explicitly included in the macro-model lead to a biased characterisation of correctly identified subhaloes as well as false positive detections.

\citet{Nightingale2022} have performed a systematic study of the degeneracies between the macro-model and isolated subhaloes. From the analysis of a sample of fifty-four gravitationally lensed galaxies from the SLACS and BELLS GALLERY surveys \citep{Bolton2008,Shu2016}, they initially reported thirty-four subhalo detections at low significance. Upon further analysis, it is found that sixteen of these thirty-four are due to a degeneracy with the lens light (see Section \ref{sec:lens_light}) and five are due to insufficient source plane resolution. The remaining thirteen false positives are macro-model dependent with eight due to an overly simple lens mass model: a subhalo is favoured because it can replicate the effect of the missing complexity. These false positives can be accounted for by including in the lens mass models more radial (e.g. the broken power-law model) and angular (e.g. multipoles) structure in two and six cases, respectively. In general, this analysis by \citet{Nightingale2022} demonstrates that \emph{no one lens is like another} and more complicated mass distributions can both remove and add a subhalo candidate depending on the lens system and the type of model considered (e.g. multipoles versus composite mass distribution). The situation is further confounded by the fact that the nature of the macro-model is unknown a priori. While stellar disks may be visually identified with appropriate observations \citep{Hsueh2016, Hsueh2017, Nierenberg2020}, this is not necessarily the case for other forms of radial and angular structure. Moreover, depending on the spatial scales affected, the latter may be identified and characterized only with very-high angular-resolution data \citep[][]{Powell2022}. These results highlight the importance of allowing for models other than a smooth SPL and modelling each lens to great detail. 

While this degeneracy is intrinsic to the measurement itself and affects any type of lens modelling technique and data, it is particularly thorny for compact sources and models that solely rely on analytical macro-models\footnote{These also include the case where low-mass haloes are described by a GRF, as the latter is an unlikely characterisation of (baryonic) structures in the lens mass distribution}. In the first case, the data (at most, eight positions and four flux measurements) provide only limited information on the lens model. In the second case, for each lens system, one needs to marginalise over a wide range of macro-model assumptions. 
Extended sources, where the potential is modelled with a free-form approach (Section \ref{sec:modelling_approaches}) have in this respect an advantage: the data has more constraining power and the freedom allowed to the potential makes it possible to detect and potentially differentiate various mass components from each other \citep[e.g.][]{Galan2022}. 

In reality, the degeneracy between the lens macro-model and dark matter fluctuations, does not act in isolation. Its net effect is the result of its interplay with another degeneracy, that between the lensing potential and the source light distribution. We discuss the latter in the following section.

\subsection{Degeneracy with the source}

\subsubsection{Compact sources}

The angular size of the background source relative to that of a low-mass halo of a given mass sets the level of perturbation to the lensed images magnification. This effect introduces a degeneracy between the amount of low-mass haloes, micro-lensing by stars in the main deflector and the unknown size of the source. Micro-lensing by stars (see Chapter 5) becomes problematic when the image flux is gathered from a region surrounding the background quasar that is less than $\sim 0.1~\rm{pc}$ in diameter. It is responsible for an overestimation of the amount of low-mass haloes and hence a bias in favour of colder dark matter models. One can mitigate or even remove its effect with observations at wavelengths that are known to originate from an extended region around the source or by monitoring the system for a sufficiently large amount of time.

Provided that the source is large enough to avoid micro-lensing, uncertainties in the intrinsic size of the emission region translates into uncertainties in the amount of low-mass haloes. An overestimation (underestimation) of the former leads to an underestimation of the latter and a bias in favour of warmer (colder) dark matter models. Typically, one tries to account for this degeneracy by adopting observationally motivated priors \citep[e.g.][]{Koopmans2003,Chiba2005,Muller2011} and marginalising over the size of the source. 

Another source of systematic error is related to flux variations in the emission region over time.
Flux changes appear with a time delay between different positions on the image plane. As a result, single-epoch measurements sample the intrinsic light curve of the source at different times for the different lensed images. This effect leads to flux measurement errors that can be as large as 20 per cent \citep{Dalal2002,Koopmans2003}. It can be mitigated by monitoring and averaging the observed flux over a long period of time, or correcting for the time-delay.

At radio wavelengths, propagation effects, such as free-free absorption and scatter broadening can alter the measured properties of the different lensed images. As these effects have a strong wavelength dependence, they can be identified and corrected for with multi-wavelength observations (e.g. Winn et al. 2004; Biggs et al. 2003; Mittal et al. 2007). 

\subsubsection{Extended sources}

The detection of low-mass haloes with extended sources is based on the idea that the many pixels on the image plane provide redundant information about the source surface brightness distribution, allowing one to separate structures in the background object (which appear two or four times in the data) and structures in the lensing potential (which produce a relatively localised effect on the lensed images). In practice, however, due to the presence of noise in the data and the smoothing effect of the telescope point spread function, a degeneracy exists. 

The extent of this degeneracy is strongly dependent on how the source and lensing potential are modelled. \citet{Vernardos2022} have studied in detail the case of a pixellated source and potential corrections. They concluded that the two can partly absorb each other's complex structures in a way that depends on the form of the regularisation, the pixellatisation resolution and the actual complexity in the data \citep[see also][]{Bayer2023a}. For example, when modelling a lens system, which is well described by a smooth analytical lens mass distribution and a complex source, the potential corrections can partly absorb fluctuations in the  latter that are not well captured by the grid resolution and regularisation. Similarly, the source can adapt to compensate for the lack of structure in the model for the lensing potential. Interestingly, \citet{Vernardos2022} find that the degeneracy between complexity in the source and in the potential mainly results in a biased lens macro-model and source, while the statistical properties of the potential corrections are well recovered. This suggests that, at least in the free-form approach, this degeneracy should not result in biased constraints on the properties of dark matter. However, a detailed investigation of this issue as a function of data type (e.g. optical or interferometric) and quality (e.g. angular resolution, signal-to-noise ratio and uv-coverage) is still lacking.

At the other end of the spectrum, models with analytical sources and potentials are likely to result in dark-matter constraints that are biased towards models that are colder or have a lower FDM particle mass. As the only complexity allowed in the analysis is in the form of low-mass haloes or FDM granules, these are then likely to absorb structures in the source as well as in the macro-model. Identifying suitable priors for the lensing potential and source light is therefore a key ingredient to infer the properties of dark matter with strong gravitational lensing.

\subsection{Degeneracy with the lens light}
\label{sec:lens_light}

Optical data includes light emission from the lens galaxy. This contribution is either pre-subtracted from the data \citep[e.g.][]{Vegetti2010b} or inferred during the lens modelling analysis \citep[e.g.][]{Ritondale2019a, Ritondale2019b, Nightingale2022}. Typically, it is described via a two dimensional B-spline function \citep[e.g.][]{Vegetti2014} or an analytical profile such as a (combination of) S\'ersic model \citep[e.g.][]{Ritondale2019b,Nightingale2022}. 
As for the lens mass distribution, simplistic models for the lens light may fail to reproduce complex radial and angular structures, as for example, boxy or disky isophotes and dust lanes. As a result, false positive detections of low-mass haloes may be obtained \citep{Nightingale2022}. It is therefore important to test candidate detections against different assumptions for the lens light \citep[e.g.][]{Vegetti2012,Nightingale2022} or, if possible, with observations at different frequencies\footnote{Ideally at frequencies at which the lens does not emit any light.}. For example, one can use observations in two or more wavelengths and the information carried by the multiple lensed images to correct for the effect of dust in the lens galaxy \citep[e.g.,][]{Suyu2009}.

\subsection{Some considerations on interferometric data}

At present, radio interferometetry \citep[e.g.][]{Smirnov2011} provides the highest angular resolution available for strong gravitational lensing observations. In principle, this makes it the most sensitive probe of the halo and subhalo mass functions, with very long baseline interferometry (VLBI) arrays capable of individual halo detections down to $10^6 M_\odot$ \citep{Mckean2015}.  However, modeling radio interferometric observations of strong gravitational lenses is challenging given the very large data sizes, and is an area of active research. 

It is preferable to model radio interferometric observations directly in the native visibility space (a visibility is a Fourier component of the sky brightness as measured by two antennas at a given time and frequencey interval). In principle, it is possible to model a gravitational lens observation by first imaging the data using some established deconvolution technique (e.g. CLEAN, \citealt{Hogbom1974}; or maximum-entropy, \citealt{Cornwell1985}), then applying standard lens modelling techniques to the deconvolved image. However, on the space of the CLEANed images, the noise is correlated and not well characterised, and the CLEANing process may introduce artefacts in the surface brightness distribution that mimic the effect of low-mass haloes. Source-plane deconvolution algorithms for radio observations of gravitational lenses (such as \rm{LensClean}; \citealt{Kochanek1992,Ellithorpe1996,Wucknitz2004}) were the first to be explored as a solution to this issue. Direct $\chi^2$ fitting in the visibility plane was then applied by several authors \citep{Bussman2012,Bussman2013,Hezaveh2013a} to observations of lensed dusty star-forming galaxies (DSFGs) taken with the Sub-Millimetre Array (SMA) and the Atacama Large Millimetre Array (ALMA). A fully Bayesian treatment for fitting interferometric observations, which included prior information on the source surface brightness, was finally realized by \citet{Rybak2015b} and \citet{Hezaveh2016b}.  

Amplitude and phase calibration errors \citep[e.g.][]{Pearson1984} in interferometric data can potentially masquerade as false-positive detections of low-mass dark matter haloes. This was investigated using ALMA observations of the DSFG SDP.81 by \cite{Hezaveh2016b}, with a test of the data sensitivity to the subhalo detection both with and without a treatment for phase errors. \cite{Hezaveh2013b} and \cite{Hezaveh2016b} addressed this issue by including a single antenna-based phase correction as a free parameter in the lens modeling, self-consistently incorporating uncertainties due to this systematic effect into the model posterior. It is expected that the problem of calibration errors is less relevant for cm-wavelength VLBI observations as the atmosphere is more stable over time and the antenna receivers are more sensitive. Strongly lensed sources observed with VLBI are typically radio-bright jets containing both extended and compact features, which help to provide robust phase calibration solutions prior to the lens modeling step. This is in contrast to mm-wavelength observations of DSFGs, which feature rather diffuse, low-surface-brightness emission; in this case modeling phase errors as part of the lens modeling pipeline likely yields more robust calibration solutions than \emph{a priori} self-calibration.  While it is clear that residual phase calibration errors have an effect on the rate of false-positive detections and sensitivity to low-mass dark matter haloes, there is yet to be a systematic study of this effect.

The size of the data is important when it comes to modeling antenna phases directly; the ALMA observation to which this phase-correction model was applied contains a small enough number of visibilities ($5\times 10^5$ after some additional binning) that a standard linear-algebra solver framework could be applied to the source inversion step \citep{Warren2003, Suyu2006}. For cm-wavelength VLBI data, for which averaging can introduce additional systematics that degrades sensitivity to low-mass dark haloes, an FFT-based iterative solver is required in order to be computationally tractable \citep{Powell2021, Powell2022}.  

\subsection{Theoretical Unknowns}
\label{sec:th_un}

\subsubsection{Resolution, halo and subhalo mass functions}

Theoretical predictions for the abundance and structure of CDM haloes and subhaloes initially came from analytical models, such as perturbation theory \citep{Zeldovich1970}, the Press-Schechter model \citep{Press1992}, the statistics of peaks in Gaussian random fields \citep{Bardeen1986} and the excursion set approach \citep{Bond1991,Sheth1999a}. Predictions from these theoretical models have then been compared to the results from N-body (i.e. dark-matter-only) numerical simulations. This comparison has led to precise fitting functions for the halo mass function that take into account the non-linear evolution of structure formation \citep{Sheth2001b,Giocoli2008a,Tinker2008,Despali2016}. Similarly, the functional form of the NFW profile \citep{Navarro1996} reflects the expectations for the density profile of a dark-matter-dominated structure that forms via hierarchical accretion. CDM N-body simulations agree with each other to the per cent level, and the impact of numerical effects and (sub) halo identification methods has been widely studied and is well understood \citep{Knebe2011,Knebe2013,Okabe2013}. 

However, some uncertainties remain. These are mainly related to the resolution limit of the numerical simulations and the consequent fact that analytical predictions have only been tested down to a finite scale. Numerical effects are especially problematic for the subhalo population. For example, the artificial disruption of subhaloes related to the limited spatial resolution of the simulations \citep{Green2019,Green2021} can lead to an underestimation of the number of such objects on scales of the resolution limit of a factor of 10 to 20 per cent. Moreover, while the density profile of isolated haloes is well understood, subhaloes are affected by tidal disruption and stripping inside the main halo and thus show a larger variety of profiles \citep{Sawala2017,Moline2017}, which are not all well described by the same functional form.  

WDM N-body simulations, in which the power spectrum cutoff is resolved, are known to undergo artificial fragmentation in filaments producing spurious clumps that, close to the resolution limit, can outnumber real structures. One challenge is thus to correctly identify and remove them from the (sub)halo catalogues: this can be achieved by studying the shapes of the Lagrangian initial regions that correspond to the final structures, and eliminating those that are very elongated \citep{Lovell2012}. Alternatively, \citet{Stucker2020} have developed a method to smooth the density field using phase-sheet methods and a high-resolution force calculation, in order to completely circumvent the issue and avoid the formation of spurious subhaloes.

FDM theories predict differences in the subhalo population of a lens, with a suppression of the halo mass function at low masses due to the large de Broglie wavelength of the dark matter particle. Characterizing the low-mass halo population in FDM cosmologies is an area of active research.  Analytic approximations to the FDM halo mass function in FDM have been attempted \citep{Marsh2014,Kulkarni2022}, but at this time only mass functions obtained from numerical simulations \citep{Schive2016,May2022} have been applied in a lens modeling context \citep{Laroche2022}.  The mass-concentration relation for FDM haloes is also highly uncertain; \citet{Laroche2022} modelled it using an extended Press-Schechter formalism \citep{Schneider2015} assuming some correspondence between the gravitational collapse time-scales of FDM and CDM haloes. Additionally, FDM haloes contain a characteristic soliton core that alters their density profiles \citep{Schive2014b,Chan2022}.  An accurate model for subhaloes around a lens galaxy is important to consider for inferences based on flux-ratio anomalies of unresolved compact sources, as subhaloes and granules can produce similar observational signatures. However, for FDM particle masses lower than $\sim 5 \times 10^{-21}~\mathrm{eV}$, subhaloes are too few and too diffuse to impart a small-scale signature on the observed source morphology, and can be absorbed into a sufficiently complex macro-model. This leaves the presence of granules as the main source of constraint on the particle mass.

\subsubsection{The effect of baryons}

In the past few years, numerical simulations, and especially those with a CDM cosmology, have made significant progress. Large-scale structure simulations are now able to reproduce realistic galaxy morphologies and the observed scaling relations \citep{Vogel2014a,Schaye2015,Pillepich2018,Dubois2021}, reducing some of the tensions between CDM and observations \citep{Brooks2013}.
Despite these successes, uncertainties, which may affect the interpretation of strong gravitational lensing studies within a given model, persist. Feedback processes, that cause a loss of baryonic mass, alter the total halo mass in a non-trivial way that depends on the halo mass and the galaxy formation model \citep{Sawala2015,Despali2017,Pillepich2018,Garrison2019}. As a result, the number density of low-mass haloes ($10^{7} M_{\odot}<M<10^{10}M_{\odot}$) is suppressed in hydrodynamical simulations by a factor between 10 and 40 per cent. This suppression is smaller than the one seen in most WDM models in the same mass range, as the inclusion of baryons further suppresses the halo and subhalo mass function \citep{Lovell2019, Despali2017}. However, due to the lack of hydrodynamical simulations with high-enough mass resolution, it is at present unclear how these mass functions are affected by the baryon and dark matter physics below the probed mass limit. 

Baryonic physics also affects the inner mass density profile of galaxies in a way that depends on the feedback model \citep[e.g.][]{Mukherjee2021} and its interplay with the dark matter model. For example, \citet{Robertson2019}, \citet{Despali2019} and \citet{Shen2022} have found that, in hydro-simulations with elastic SIDM, haloes have a larger variety of density profiles (with respect to the non-elastic SIDM case) and that the shape of haloes is much closer to the CDM hydrodynamical case, than initially inferred from dark-matter-only simulations. \citet{Mocz2019} studied the interplay between baryons and fuzzy dark matter, finding that the first stars can form in filamentary structures along the cosmic web, instead of only in the collapsed haloes. These results, could have interesting implication for the analysis of strong gravitational lens galaxies. However, the limited spatial resolution available in simulations at the scale of massive galaxies prevents us from a robust comparison between observations and simulations. For example, simulated galaxies display a central density core, which is large enough to produce strongly lensed image configurations that are not observed. While these uncertainties affect simulations in all dark models, they are currently more problematic for alternative non-CDM models. The vast majority of simulations where dark matter is different from CDM does not include baryonic physics \citep{Lovell2012,Vogel2014b} and is not yet at the level of the CDM case, in terms of resolution, number of objects and volume \citep{Adhikari2022}. Moreover, even when baryons are included, the sub-grid physics processes that describe their behaviour often remain calibrated on CDM simulations. Hence, it is  unclear whether non-CDM hydro-dynamical simulations provide a correct description of the interplay between dark-matter and baryons. 
Obtaining large samples of numerical simulations with different galaxy formation and dark matter physics, as well as the volume and resolution required for a meaningful comparison with strong gravitational lensing observations is a fundamental step for future works.

\section{Historical perspective}
\label{sec:history}

\subsection{Compact sources}

The use of strong gravitational lensing as a probe of dark matter was first proposed by \citet{Mao1998}. They showed that anomalies in the flux ratios (i.e. ratios that deviated from those predicted by smooth mass models) of lensed quasars could be accounted for by subhaloes in the lens galaxy. In particular, they focused on the two and three
brightest images in a fold and cusp configuration, respectively. For a smooth lens mass distribution, the fluxes of these lensed quasar images should satisfy the asymptotic relations in Eqs. (\ref{equ:fold}) and (\ref{equ:cusp}). Discrepancies between the observed flux ratios and the generic predictions are therefore an indication of some type of small-scale structure in the mass distribution.
\citet{Mao1998} also pointed out that flux-ratio anomalies observed in optical data could easily be due to micro-lensing by stars in the lensing galaxy. Focusing on radio observations was, therefore, a better way to investigate the presence of subhaloes. At these frequencies, the lensed emission had a large enough angular extent as to be insensitive to micro-lensing by stars \citep[although see ][]{Koopmans2000}. The paper investigated both compact mass distributions (globular clusters with masses of $\sim 10^6 M_\odot$) and smoother fluctuations such as spiral arms in the lensing galaxy.  

\citet{Mao1998} thus truly set the stage for subsequent investigations of perturbations by dark matter haloes by: (1) considering perturbations by larger-scale structures than stars and, (2) pointing out that the observations had to be conducted using sources that had angular scales that were large compared to the Einstein ring radii of the stars in the lensing galaxy.  This second requirement ruled out using emission from the accretion disks and broad-line regions associated with the lensed AGN, which are sensitive to micro-lensing. However, observations at radio or mid-IR wavelengths could be used to investigate dark matter since the regions in the background objects that produce emission at these wavelengths are large enough to be mostly unaffected by micro-lensing by stars in the lensing galaxy, but small enough to be sensitive to perturbation by relatively low-mass dark matter haloes.  Unfortunately, it was difficult to obtain large samples at these wavelengths, since only $\sim$10 per cent of AGN are radio loud, and ground-based observations at mid-IR wavelengths are extraordinarily difficult due to thermal emission from the Earth's atmosphere.  Thus, for many years the sample of lenses that were useful for dark matter investigations was on the order of 10 systems, primarily discovered in the MG \citep{MGsurvey1,MGsurvey2,MGsurvey3,MGsurvey4}, JVAS \citep{jvas1,jvas2,jvas3}, and CLASS \citep{class1,class2} radio surveys.  
Initial investigations focused on individual lenses, particularly radio-loud four-image systems that strongly violated the standard relationships for merging images \citep{Fassnacht1999,Marlow1999,Trotter2000,Biggs2004}, and included some work that incorporated numerical simulations to understand subhaloes \citep{Bradac2002}.

The seminal paper by \citet{Dalal2002} was the first analysis of flux-ratio measurements in a statistically significant sample, with a goal of testing the $\rm{\Lambda}$CDM model.  They examined the flux ratios in a sample of seven lens systems and found broad consistency with CDM.  Although this analysis had a high impact for many years, it did have several shortcomings in the context of current approaches to using flux-ratio statistics to draw inferences on the nature of dark matter. These include using CDM-only simulations, using fairly simple lens models, only considering subhaloes within the halo of the primary lensing galaxy while not including line-of-sight haloes, using PJ mass profiles for the subhaloes, and being restricted to using the somewhat uncertain flux-ratio measurements that were available at the time.  

For nearly two decades following this analysis, no new lens systems that had high-sensitivity radio or mid-IR flux ratio measurements were discovered, so subsequent investigations had to focus on extending or improving the analysis rather than working with larger samples.  One improvement came from a monitoring program to look for extrinsic variability in radio-loud lenses, which had the additional benefit of providing high-precision flux-ratio measurements after correcting for any time delays in the systems  \citep{Koopmans2003}.  Other work investigated astrometric shifts and parity dependence as methods for determining the abundance of subhaloes \citep{Chen2007,Chen2009}. 

Another major effort addressed the problem by conducting ray-tracing analyses through high-resolution numerical simulations, which allowed the inclusion of lower subhalo masses and, thus, more thoroughly to explore the lensing effects of dark-matter subhaloes and other possible perturbers \citep{Mao2004,Xu2009,Xu2010,Richardson2022}. These studies incorporated not only the number of anomalies in the observed radio lenses, but also how much the observed flux ratios deviated from the predictions of smooth mass models.  

Interestingly, several of these studies found that CDM subhaloes were not sufficient to explain the observations fully \citep{Xu2009,Xu2015}, except maybe when including line-of-sight haloes \citep{Xu2012}, suggesting that more complex macro models were needed.  Earlier work had reached a related conclusion from simulated data, finding that an edge-on stellar disk could cause violations of the standard cusp relation for the lensed image magnifications \citep{Bradac2004}.  Subsequent work discovered exactly this type of situation in observed lens systems.  In some systems with the most extreme flux-ratio anomalies, high-resolution imaging from ground-based adaptive optics and HST data revealed edge-on disk components of the lensing galaxies. These additional baryonic components could explain the observed flux-ratio anomalies without needing to resort to either subhaloes or line-of-sight haloes \citep{Hsueh2016,Hsueh2017}. Further investigations used simulated galaxies to confirm the importance of baryonic structures in the lensing galaxies \citep{Gilman2017,Hsueh2018}.

In parallel with the work on extending and improving the analyses was an effort to increase the sample size of observed four-image lenses by pushing past the traditional radio and mid-IR observations that had provided the primary samples for the analyses above.  In particular, \citet{Moustakas2003} proposed a spectroscopic technique based on the differential magnification of several different emitting regions in the lensed quasar, i.e., the continuum, broad-line region, and narrow-line region (NLR). This approach allowed them to separate contributions from the smooth lens model, micro-lensing, and lensing by subhaloes. This method was applied to a single very low-redshift lens that was observed with an integral field unit (IFU) spectrograph, and seemed promising \citep{Metcalf2004}.  It is difficult to apply this technique to higher redshift systems using the seeing-limited ground-based observations used in \cite{Metcalf2004}, due to angular resolution issues.  However, a similar approach has been successful when using either ground-based adaptive optics IFUs \citep{Nierenberg2014} or HST grism observations \citep{Nierenberg2017}. This technique has led to the first major increase in the sample size of four-image systems with measurements that are useful for flux-ratio investigations \citep{Nierenberg2020}.

\subsection{Extended sources}

\citet{Koopmans2005} was the first to propose the use of galaxy-galaxy lensing observations to detect subhaloes within the lens using the semi-linear pixellated approach described in Section \ref{sec:modelling_approaches}. A few years later, \citet{Vegetti2009a} extended the original idea to be fully embedded within the context of Bayesian statistics. 

Then, \citet{Vegetti2009b} proposed a statistical framework for the interpretation of both detections and non detections within the context of specific theoretical predictions. They also showed how the achievable level of constraints on the subhalo mass function parameters is related to the number of lenses in the sample and the sensitivity of the data. Recently, \citet{Despali2022} have provided a detailed quantification of how the latter depends on the data properties. We discuss their findings and relative implications in Section \ref{sec:current}.

\citet{Vegetti2010b} and \citet{Vegetti2012} were the first one to apply the pixellated gravitational imaging technique to real data taken with the HST and Keck-AO. They reported a 16-$\sigma$ and 12-$\sigma$ detection of two individual subhaloes in the gravitational lens systems SDSSJ0946+1006 (also known as the double ring or the Jackpot) and B1938+666, respectively. These detections, obtained in a pixellated fashion, were then modelled with a PJ profile. From the latter, a total mass of $\sim 10^9 M_\odot$ and $\sim10^8 M_\odot$ was inferred for each system, respectively. Both detections have been independently confirmed \citep{Minor2021,Sengul2021}. However, \citet{Despali2018} and \citet{Sengul2021} have found that the detection in the system B1938+666 is more likely a field halo. Whether the concentration of these two objects is consistent with CDM predictions is currently under investigation \citep[e.g][]{Minor2021, Sengul2022}. From the analysis of the ALMA long baseline campaign data for the lens system SDP.81, \citet{Hezaveh2016b} reported the detection at the 5-$\sigma$ level a subhalo with a total PJ mass of $\sim 10^9 M_\odot$. \citet{Inoue2016} also found a detection in this lens system. However, the inferred subhalo position and lens macro-model are inconsistent between the two analyses. Further investigations are required to understand the origin of this discrepancy. 

Focusing, for the first time, on a larger number of strong gravitational lens systems, \citet{Vegetti2014} found no additional subhaloes in a sample of ten SLACS lenses, in the mass regime probed by the data. Recently, \citet{Nightingale2022} have searched for the presence of subhaloes in fifty-four lens systems (the largest number considered so far) from the SLACS and BELLS gallery samples. They reported two candidate detections, one of which matches the one by \citet{Vegetti2010b} in  the Jackpot lens. 

While early studies considered the subhalo population only, \citet{Li2017}, \citet{Despali2018} and \citet{Amorisco2022} have shown the contribution from field haloes to be important and in some cases dominant. This is a significant result. Unlike for subhaloes the properties and number of field haloes are better understood from a theoretical perspective (see Section \ref{sec:th_un}), and the resulting increase in the number of detectable objects per lens allows for stronger constraints on dark matter with fewer systems. 
Allowing for the contributions of both populations, \citet{Ritondale2019b} concluded that their lack of detection in twenty-one lens systems from the  HST BELLS gallery sample is consistent with the CDM model.

Dark-matter constraints from strongly lensed extended sources have been mainly limited by the amount of available data with enough angular resolution. In recent years, efforts have been made to increase the number of known strong gravitational lens systems \citep[e.g.][]{Lanusse2018,Huang2021,Petrillo2019b,Canameras2021,Rezaei2022}, as well as the number of observations with improved angular resolution with, for example, Keck-AO \citep{Lagattuta2012} in the NIR, ALMA in the sub-mm \citep{Spilker2016} and VLBI at cm-wavelenght \citep{Spingola2019}. At same time, several new lens modelling approaches have been developed (see Section \ref{sec:modelling_approaches} and references therein).

Existing data-sets as well as simulated ones, have been used to improve our knowledge of systematic errors and degeneracies  (see Section \ref{sec:dsu} and references therein), and our  understanding of the signal under study. For example, \citet{Amorisco2022} showed that an improvement on the level of dark matter constraints can be obtained by taking into account the low-mass haloes mass-concentration relation (and how it changes with the dark-matter model) and its scatter. Similarly, ray-tracing through numerical simulations has been instrumental to quantify the lensing effect arising from smaller- and larger-scale structures \citep[e.g.][]{Enzi2020,He2022} in different dark matter models \citep{Despali2019,Despali2020}.

\section{Dark matter constraints}
\label{sec:current}

\begin{figure*}
\begin{center}
	\includegraphics[width=0.8\textwidth]{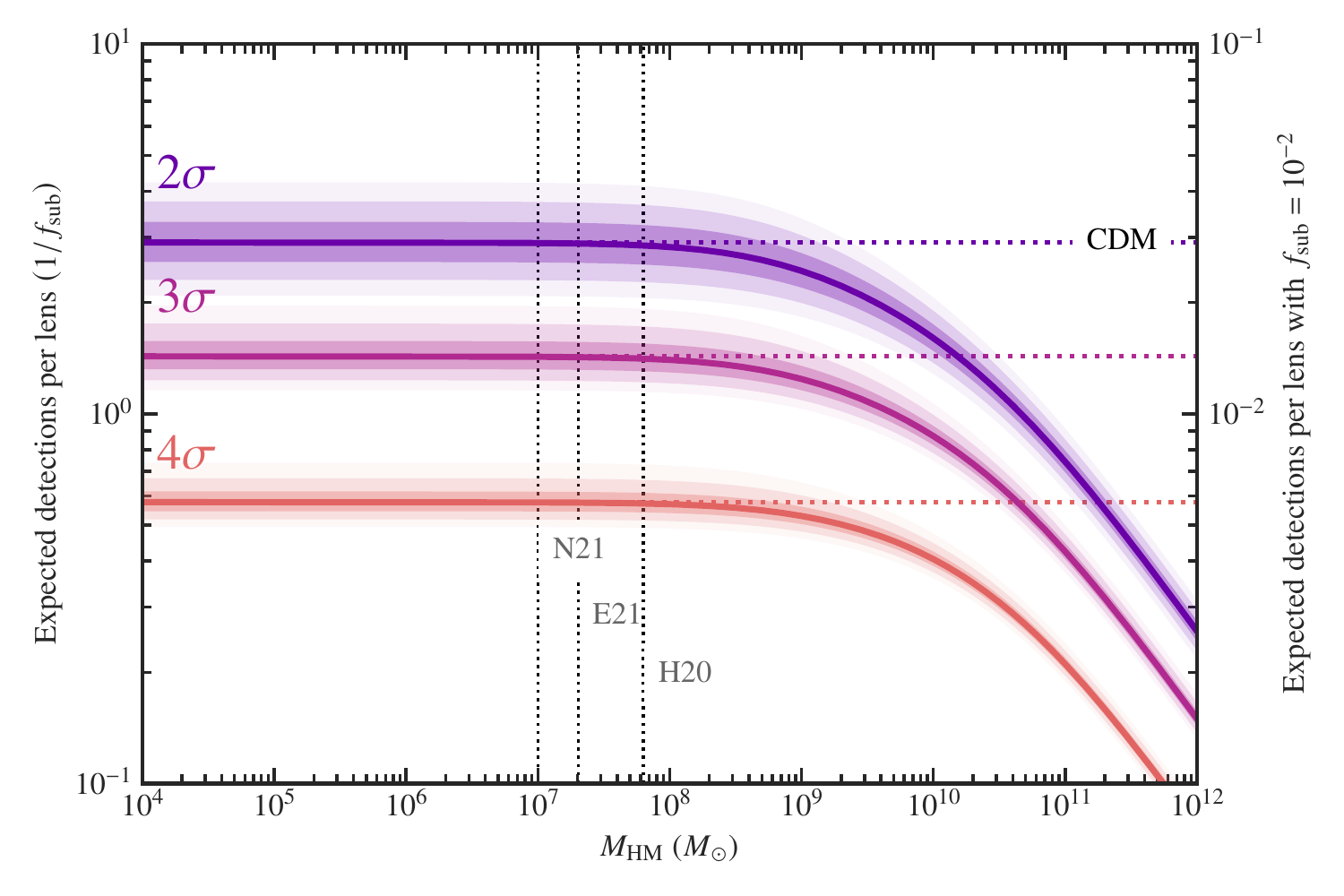}
\end{center}	
 \caption{\label{fig:wdm_constraints}
	Expected number of subhaloes per lens as a function of the half-mode mass from Euclid-like observations. Each curve (with 64 per cent, 95 per cent, and 99 per cent confidence areas) is for a different level of significance of subhalo detection. The horizontal dotted lines display the expected number of detectable subhaloes in CDM \citep[see][for more details]{ORiordan2023}. The vertical dotted lines show the current 95 per cent upper limits on the half-mode mass from \citet{Nadler2021b} and \citet{Hsueh2020} together with the 1/20th of the maximum Likelihood from \citet{Enzi2021}.} 
\end{figure*}

\subsection{Compact sources}

Since 2019, four new analyses \citep{Hsueh2020,Gilman2020a,Gilman2020b,Laroche2022} have characterized the properties of low-mass haloes using a sample of quadruply-imaged quasars. These works interpret strong lensing data in the context of a variety of scenarios, including classes of dark matter such as warm dark matter, ultra-light dark matter, and self-interacting dark matter, as well as early Universe physics that alters the power spectrum of primordial density fluctuations. 
 
\citet{Hsueh2020} and \citet{Gilman2020a} analyzed a sample of strong lens systems in the context of warm dark matter, presenting evidence for an unseen population of dark matter haloes with masses around $10^8 M_{\odot}$. This inference places an upper limit on the free-streaming length of dark matter equivalent to that of a 5 to 6 $\rm{keV}$ thermal relic dark matter particle, and constrains a variety of sterile neutrino models with varying production mechanisms \citep{Zelko2022}. \citet{Gilman2020b} used the sensitivity of the relative magnifications among the lensed images to infer the concentrations of CDM haloes, an analysis that was later generalized by \citet{Gilman2022a} to make a direct connection between the abundance and internal structure of dark matter haloes and the primordial matter power spectrum on small scales $\left(k > 10 \ \rm{Mpc^{-1}}\right)$. The resulting inference on the concentration-mass relation and power spectrum agreed with the CDM prediction. \citet{Laroche2022} interpreted the same sample of eleven four-image lenses as analysed by \citet{Gilman2022a} in the context of ultra-light dark matter, showing that the granular structure of the host halo density profile that arises from quantum wave interference effects can impact image flux ratios in a similar manner to dark matter haloes.  Finally, building on work by \citet{Gilman2021}, \citet{Gilman2022b} interpreted the same sample of eleven lenses in the context of self-interacting dark matter and showed that existing data disfavors SIDM models with large amplitudes at low speeds, such as those that can arise from resonances in the self-interaction cross section, assuming the large amplitude of self-interaction cross section drives low-mass haloes to core-collapse. By combining observations of the Milky-Way satellites with the sample of strongly lensed compact sourses from \citet{Gilman2022a}, \citet{Nadler2021b} have derived a limit on the half-mode mass of $M_{\rm hm}<10^7 M_\odot$ (i.e., $m_{\rm WDM}$ > 9.7 keV) at the 95 per cent confidence level (see Fig. \ref{fig:wdm_constraints}).

These analyses have incorporated the latest theoretical understanding of structure formation in CDM and alternative dark matter models, such as the halo mass function and concentation-mass relation in warm dark matter, the process of core-collapse that is expected to occur in self-interacting dark matter, and the phenomenon of wave-interference unique to ultra-light dark matter. At the same time, \citet{Basu2021} have introduced a novel approach to constrain Axion-Like Particles dark matter models from the differential birefringence effect imparted on the strongly lensed images. Using broad-band polarisation observations of the lens system B1152+199 from the Karl G. Jansky Very Large Array (VLA), they derived an upper bound on the ALP-photon coupling between $g_{\rm{a \gamma}} \leq 9.2 \times 10^{-11}$ and $7.7 \times 10^{-8}$ eV at the 95 per cent confidence limit for an ALP mass between $m_{\rm{a}} = 3.6 \times 10^{-21}$ eV and $4.6 \times 10^{-18}$ eV.

Two recent innovations have expanded the scope of strong lensing of compact sources as a probe of fundamental dark matter physics. First, the sample size of lenses suitable for a subhalo inference doubled with measurements of relative image fluxes from narrow-line emission around the background quasar. Nuclear narrow-line emission, which emanates from an area in the source plane with a typical size between 10 to 100 $\rm{pc}$, subtends angular scales on the sky much larger than a micro-arcsecond. This renders the relative brightness of lensed images immune to micro-lensing, while retaining sensitivity to milli-lensing by dark matter haloes. Second, open-source software to model gravitational lens systems and to generate populations of dark matter haloes for lensing computations have enabled ray-tracing simulations to be performed in parallel on computing clusters with realistic background sources and full populations of subhaloes. At the same time, recent analyses (from 2021 onwards) have made some efforts to mitigate potential systematic uncertainties associated with the lens macro-model by including the contribution of multipole mass moments.

\subsection{Extended sources}

\begin{figure}
	\includegraphics[width=8.5cm]{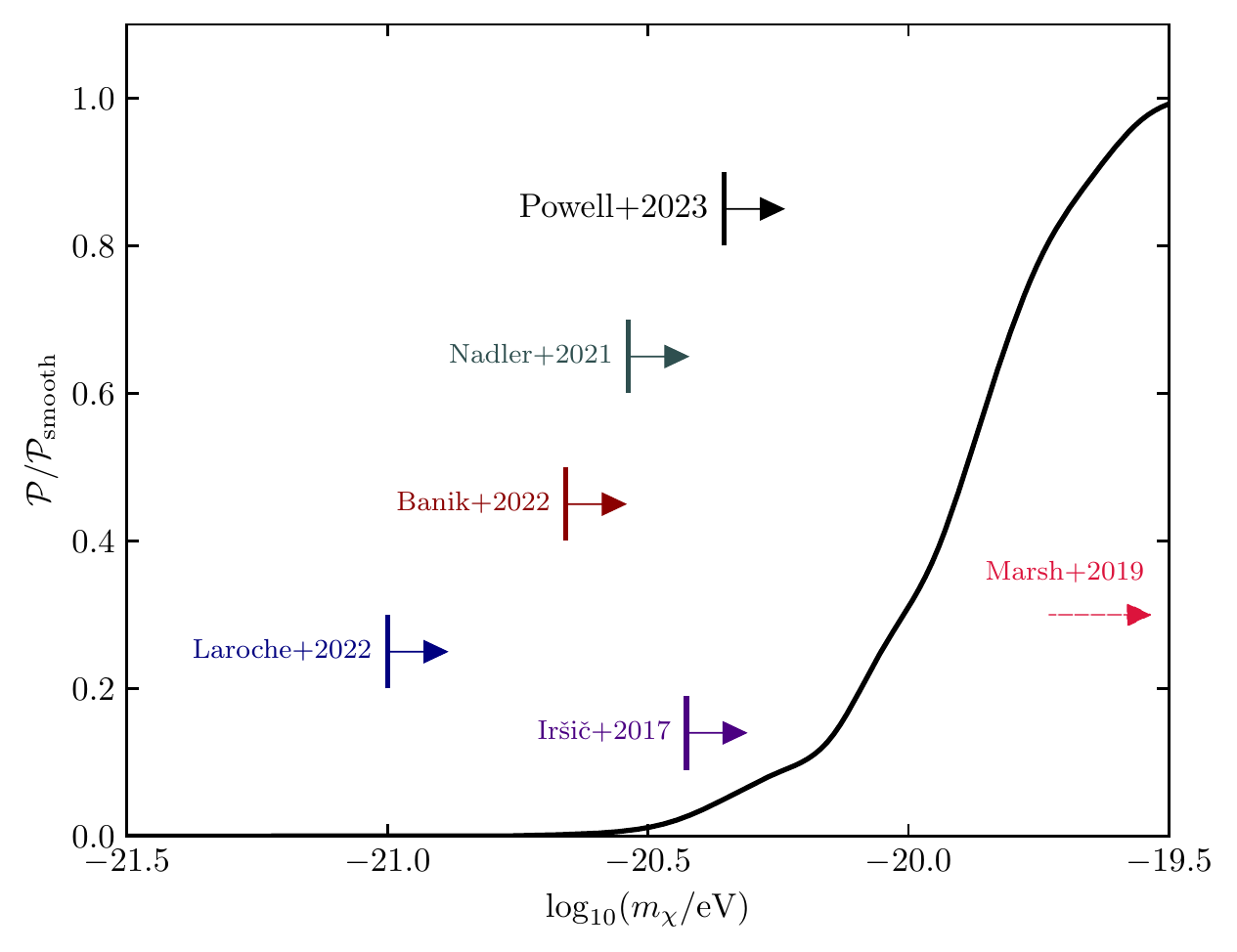}
	\caption{\label{fig:FDM_constraints}
	Bounds on the FDM particle mass from different observational probes.
    The solid black curve and arrow show the fully marginalized posterior odds ratio and relative lower bound from \citet{Powell2023}, respectively. These were derived from the analysis of mas-resolution observations of the lens system MG J0751+2716 taken with the Global Very Long Baseline Interferometry array. The results by \citet[][blue arrow]{Laroche2022} were obtained from a sample of eleven strongly lensed QSOs. Also shown are the constraints from the Lyman-$\alpha$ forest \citep[][purple arrow]{Irsic2017}, the Milky-Way satellites \citep[][green arrow]{Nadler2021a}, stellar streams in the Milky-Way \citep[][dark red arrow]{Banik2021}. The limit by \citet[][red dashed arrow]{Marsh2019} lie beyond the limit of the plot, and were derived from observations of star clusters within the Milky-Way Ultra-faint Dwarf satellite galaxy Eridanus II.} 
\end{figure}

Most studies based on extended lensed sources are so far consistent with the CDM model. For example, \citet{Vegetti2014} and \citet{Hezaveh2016b} have inferred a dark matter fraction in subhaloes that is consistent with the subhalo population in CDM simulations \citep{Springel2008,Xu2015,Despali2017}. From an analysis (which only include the contribution of subhaloes) of the lens system RXJ1131$-$1231, \citet{Birrer2017} have derived a 2-$\sigma$ lower limit on the particle of a thermal relic dark matter model of $m_{\rm th} > 2$ keV. On the other hand, \citet{Bayer2023b} have found an upper limit on the power-spectrum of mass density fluctuations in the lens system SDSS J0252+0039 that exceeds the value expected from CDM. Their result is likely driven by their choice of macro-model (SPL) and the lack of other form complexity beyond subhaloes.

Allowing for the contribution of field haloes, \citet{Vegetti2018} have derived constraints on sterile neutrinos that show a preference for colder dark matter models: $\log {M_{\rm hm}} [M_\odot] < 12.0$  at the 2-$\sigma$ level. This result excludes sterile neutrino models with neutrino masses $m_{\rm s} < 0.8$ keV at any value of the lepton asymmetry $L_6$. \citet{Ritondale2019b} reported zero detection of low-mass haloes in a sample of twenty-one strong gravitational lens systems from the BELLS survery, in agreement with CDM given the quality of the data. Recently, \citet{Enzi2021} have derived constraints on thermal relic and sterile neutrino dark matter models by combining the lensing results by \citet{Vegetti2018} and \cite{Ritondale2019b} with observations of the Lyman-$\alpha$ forest and the Milky-Way satellite galaxies. They derived a joint limit on the thermal relic mass of $m_{\rm th} > 6.048$ keV (i.e. $M_{\rm hm} < 3 \times 10^7 M_\odot h^{-1}$) at the 95 per cent confidence level (see Fig. \ref{fig:wdm_constraints}). This result is mainly set by the Milky-Way and the Lyman-$\alpha$ forest. The lensing measurements alone lead to a 95 per cent confidence level lower limit of $m_{\rm th} > 0.6$ keV and $m_{\rm th} > 0.1$ keV, for the SLACS and BELLS samples, respectively. 

While most published analyses of extended sources show a preference for CDM, they do not rule out alternative and still viable dark matter models yet. This lack of constraints is partly due to the low number of lens systems available and the relatively limited quality of the data: signal-to-noise ratio and in particular angular resolution. Using realistic mock observations of varying data quality, \citet{Despali2022} have recently quantified how the sensitivity to subhaloes depends on the properties of the lens system (e.g. source structure and position relative to the caustics) and the data (i.e. signal-to-noise ratio and angular resolution). They concluded that the lowest detectable subhalo mass decreases linearly with signal-to-noise ratio, and more strongly with the angular resolution. An increase in the latter by a factor of two leads on average to an increase in mass sensitivity of a similar factor (for a fixed signal-to-noise ratio, lensing configuration and source properties). For example, existing HST observations ($<\rm{SNR}>\sim 3.5$ and PSF$_{\rm FWHM}=0.09$ mas) have a maximum sensitivity of $M_{\rm vir}^{\rm NFW} = 10^{9}M_\odot$, leading to an expected number of detectable subhaloes per lens in CDM consistent with the current number of objects detected so far \citep{Vegetti2014,Ritondale2019b,Nightingale2022}. These effects also explain why the constraints are, at present, less stringent (though possibly more robust) than those obtained with compact sources: predictions from different dark matter models mostly differ at halo masses smaller than currently probed.

Observations at very high-angular resolution are therefore key to probe the nature of dark matter with extended sources. For example, from observations of the gravitational lens system MG J0751+2716 taken with the Global Very Long Baseline Interferometry array, \citet{Powell2023} have ruled out scalar FDM models with a particle mass $m_{\chi} \leq 4.4 \times 10^{-21}$ eV. Their constraints, which were obtained from a single lens system with milli-arcsecond angular resolution, are more stringent than previously published results from larger samples of lensed quasars (see Fig. \ref{fig:FDM_constraints}). Their work, together with the results by \citet{Despali2022} and \citet{ORiordan2023}, clearly demonstrates how a smaller number of high-angular resolution observations can be more effective at constraining the properties of dark matter than many lens systems with lower quality data. We will discuss this further in the context of future surveys in the following section.

\section{Future prospects}
\label{sec:future}

At present, the level of constraints that can be obtained on the properties of dark matter from both extended and compact sources has mainly been limited by the amount of available data. Strong gravitational lensing is a rare phenomenon and the number of known systems with suitable data (e.g. high enough signal-to-noise ratio and angular resolution for extended sources and the right frequency coverage for compact sources) is small (of the order of a few tens).

Thanks to ongoing wide-sky surveys the number of known strong gravitational lens system has significantly increased in recent years \citep[e.g.][]{Storfer2022}. With the advent of the next generation of sky surveys with, e.g., Euclid, the Low-Frequency Array (LOFAR), the Square Kilometre Array (SKA) and the Vera Rubin Observatory, this number is expected to increase even further \citep[e.g.][]{Oguri2010,Mckean2015}. The size of these samples is unprecedented. Coupled with follow-up observations they will provide robust and meaningful constraints on dark matter.

\subsection{Compact sources}

Increasing the sample size for four-image systems will reduce the statistical uncertainties of measurements made with these types of system. Provided that all sources of systematic error can be accounted for, this increased constraining power can lead to stronger constraints over various dark matter models,  such as warm and self-interacting dark matter \citep{Gilman2019,Gilman2021,Hsueh2020}. The existing analysis techniques applied to interpret data from multiply-imaged quasars can scale to match the increased sample size. Thus, computational costs will not limit the scientific output achievable with forthcoming data. 

Follow-up observations will be required to precisely measure the positions and relative magnifications among the unresolved images of the lensed source. These observations can be performed with spaced-based observatories, such as HST and JWST (while they remain operational), but most likely with ground-based facilities with adaptive optics for optical and infrared-bright quasars \citep{Nierenberg2014}, and with the Very Large Array (VLA), the enhanced Multi Element Remotely Linked Interferometer Network (e-MERLIN) and VLBI for compact lensed radio sources. Upcoming observational facilities, such as the Keck All sky Precision Adaptive optics program \citep{Wizinowich2022} and thirty-metre class telescopes, like the European Extremely Large Telescope (ELT), will provide exquisite astrometry and sensitivity with which to measure the relative image fluxes and positions. Likewise, the SKA will provide high resolution imaging at a resolution of 30 to 70 mas, which, when combined with VLBI, can provide astrometric measurements at sub-mas-arcsecond precision (e.g. \citealt{Spingola2020}).

The JWST will soon deliver precise measurements of relative image fluxes in the mid-infrared at $>$20$~\mu \rm{m}$ \citep{Nierenberg2021}, emission that emanates from a more compact region (1 to 10~pc) around the background source than the nuclear narrow-line emission (10 to 100~pc) that has been measured with the HST and Keck-AO \citep{Nierenberg2020}. The spatial extent of the mid-IR emission renders these data immune to stellar micro-lensing and variability of the background quasar, systematic effects that one could conflate with the perturbation by low-mass haloes \footnote{Scattering of light into our field of view by the spatially-extended nuclear narrow-line and mid-IR emission regions acts as a low-pass filter that washes out variability in the quasar light curves on timescales less than the light crossing time.}. The more compact mid-IR emission that is measurable with the JWST increases sensitivity to perturbations by low-mass haloes relative to existing data because the minimum deflection angle that can affect the data scales with the angular size of the lensed source \citep{Dobler2006}. Mid-IR flux ratios measured with the JWST have anticipated sensitivity to populations of haloes less massive than $10^7~M_{\odot}$ \citep{Nierenberg2021}. Moreover, the compact background sources measurable with JWST will increase sensitivity to the internal structure of low-mass haloes. This improved sensitivity, in combination with the expanded sample size of strong lens systems, will enable searches for populations of core-collapsed haloes, such as structures with logarithmic central density profile slopes steeper than $-2$ \citep{Balberg2002,Turner2021}. If detected, the existence of these objects would suggest dark matter has a velocity-dependent cross section with strong self-interactions at low speeds that triggers core collapse \citep{Gilman2021,Turner2021,Yang2022,Gilman2022b}. Increased sensitivity to the internal structure of low-mass haloes will also enable stronger constraints on the concentration-mass relation of low-mass haloes \citep{Gilman2020b}. This type of measurement can be interpreted in the context of the primordial matter power spectrum on scales $k > 10 \ \rm{Mpc^{-1}}$ \citep{Zentner2003,Gilman2022a}. However, as the background sources will remain unresolved, their unknown surface brightness distribution will still stand as a potential source of systematic error. Moreover, in  those cases without any extended emission from the quasar-host galaxy, the limited number of observational constraints provided by the data may lead to a potential systematic error in the macro-model.

At radio wavelengths (between around 4 and 18 GHz), large fractional bandwidths with, for example, the ngVLA and the SKA, will detect the emission from pc-scale radio sources, which are expected to be immune from micro-lensing (but see, e.g., \citealt{Koopmans2000,Biggs2023}). The orders of magnitude improvement in sensitivity will allow monitoring of the lensed radio sources for cosmology (in the case of any intrinsic variability) or for dark-matter studies, when the radio source is found not to vary. While the large fractional bandwidths will allow radio propagation effects, such as free-free absorption or scattering, to be identified, which can either corrected for or used to create samples of lensed radio sources that do not show such issues. However, the main contribution that radio-selected samples can provide is our ability to characterise the source structure on mas-scales through observations with VLBI \citep[e.g][]{Mckean2007,Spingola2020}. In general, it is worth noting that combining the data from multiple emission regions (mid-IR-bright torus, NRL, radio-jets) will allow various systematics associated with the source model to be quantified and possibly corrected for. This may include better constraints to the macro-model via large scale source properties, such as the optical/infrared emission from the stellar component of the quasar host galaxy or through high resolution imaging of the kpc-scale dust emission with ALMA (e.g. \citealt{Stacey2021}).

\begin{figure}
	\includegraphics[width=8.5cm]{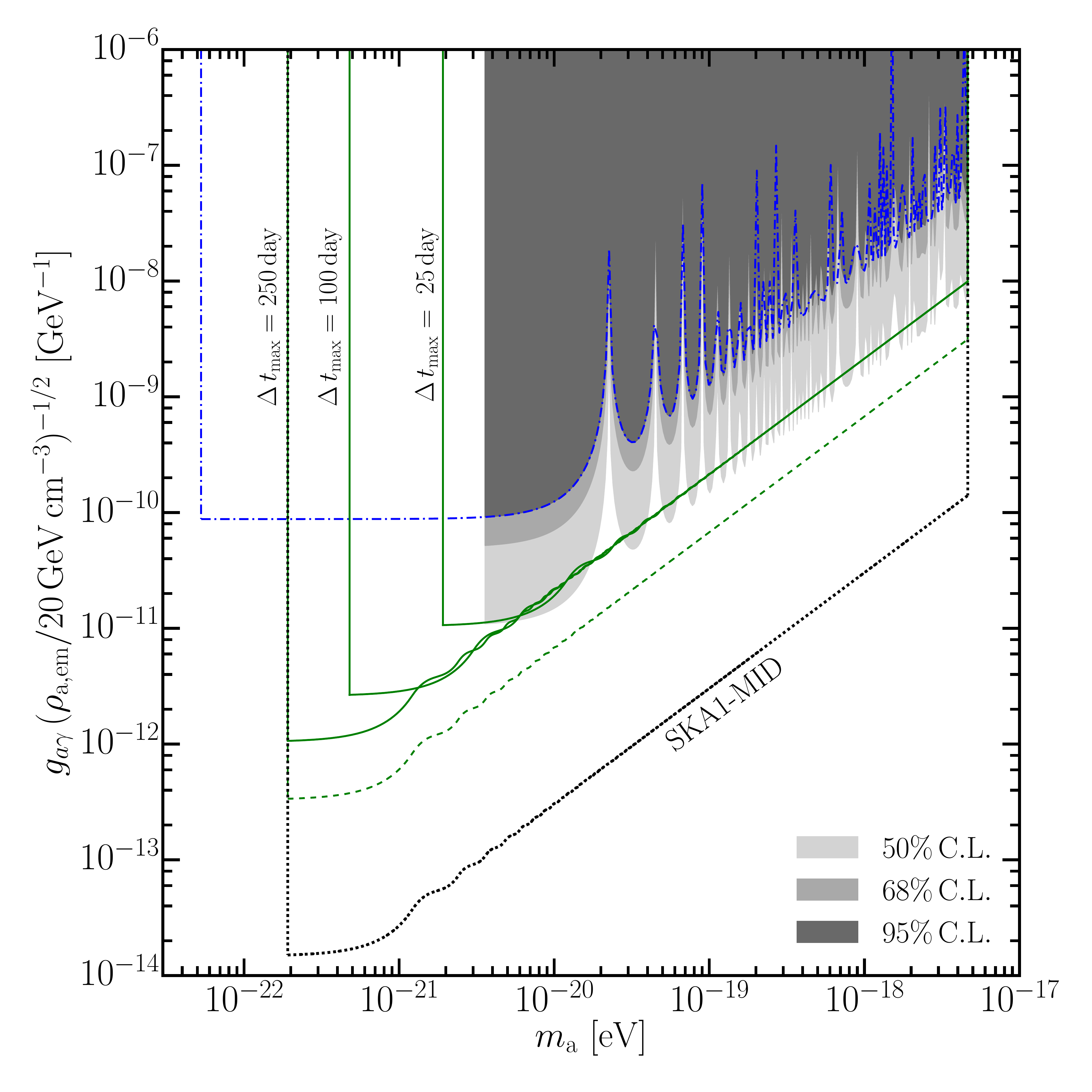}
	\caption{\label{fig:Axions}
	Limits on the ALP-photon coupling from strong gravitational lensing.
    The dashed area shows current bounds from the gravitational lens system B1152+199. The dash-dotted blue line shows the expected constraint that can be obtained from monitoring B1152+199 over the course of $\sim 5$ year. The solid and dashed green lines show the predicted bounds for a sample of 100 and 1000 gravitational lens systems for different values of the  maximum time delay ($\Delta\,t_{\rm max}$) on the observed plane. The dotted black line shows the parameter space that can be probed with the SKA. The figure is taken from \citet{Basu2021}.} 
\end{figure}

Finally, high-sensitivity observations with the LOFAR2.0 and SKA-MID of large samples of polarised and strongly lensed sources are expected to improve upon current bounds on the ALP-photon coupling by one to two orders of magnitude, depending on the ALP mass range \citep[see][and Fig. \ref{fig:Axions}]{Basu2021}. Providing, thereby, competitive constraints on dark-matter models made of ALPs.

\subsection{Extended sources}

The larger number of strong gravitational lens systems with an extended source that will be discovered by wide-sky surveys will represent a unique opportunity. However, follow-up observations at high angular resolution will be necessary to obtain meaningful constraints on the properties of dark matter. Recently, \citet{ORiordan2023} calculated the sensitivity function for 16000 Euclid-VIS-like mock observations (see Fig. \ref{fig:wdm_constraints}). They found most of the lenses in the sample to be completely insensitive to the presence of subhaloes with a mass lower than $M_{\rm max}^{\rm NFW} = 10^{11} M_\odot$. The most sensitive pixels yield a lowest detectable mass of $M_{\rm max}^{\rm NFW} = 10^{8.8\pm0.2} M_\odot$. Assuming CDM and a dark matter fraction in subhaloes of 0.01, this sensitivity leads to one detectable object in every seventy lenses. These results are in line with \citet{Despali2022} and the current number of detected objects (see previous section). Hence, Euclid will allow one to detect a number of subhaloes that is significantly larger than what has been possible so far, and potentially deliver tighter constraints on the amount of dark matter in subhaloes (i.e. the normalisation of the subhalo mass function) than currently possible. However, most of these objects will be relatively massive, and will probe (thermal relic) dark matter models with a half-mode mass of $M_{\rm hm} >10^8 M_{\odot}$. These models have already been ruled out by other observations, including strongly lensed quasars. Similarly, if not slightly worse results are expected from the Vera Rubin Observatory, given its slightly worse angular resolution.

While the sensitivity to the presence of low-mass haloes increases linearly with decreasing signal-to-noise ratio, the angular resolution of the observation sets a hard limit on the lowest detectable halo and the spatial scales on which the dark matter distribution can be probed \citep{Despali2022,ORiordan2023}. For this reason, follow-up observations with Keck-AO, HST and ALMA are likely to lead to an increase in sensitivity (relatively to Euclid) between two to three orders in magnitude in halo mass. With an angular resolution of a few milli-arcseconds, Global VLBI observations at 1.6 to 15 GHz, will allow one to detect masses as low as $10^6M_\odot$ \citep{Mckean2015}, and probe the general spatial distribution of dark matter on sub-kpc scales \citep[e.g.][]{Powell2023}. Moreover, searches for milli-arcsecond scale separation images with VLBI will potentially reveal the presence of super-critical low-mass haloes, which existence is predicted by certain SIDM models \citep{Casadio2021,Loudas2022}. In the future, 30-metre class optical telescopes such as the ELT are likely to provide a similar level of sensitivity, while interferometric observations in the 10 to 100 GHz range with SKA-VLBI or the ngVLA will lead to an even stronger increase. This significant improvement in data quality will also result in better constraints on the properties of the lens macro-model. This effect will significantly reduce the role of one of the main sources of systematic uncertainty and will potentially have interesting implications for other scientific applications of strong gravitational lensing. A relatively small number of high-resolution observations is, therefore, expected to potentially deliver stronger constraints on the properties of dark matter than the several orders of magnitude larger sample of Euclid and Vera Rubin lenses. Note, however, that these facilities and also LOFAR and the SKA will be vital for finding the lenses needed for high-resolution follow-up observations. 

\section{Summary and conclusions} 
\label{sec:conclusions}

Strong gravitational lensing provides a unique channel to constrain the dark matter distribution on subgalactic scales, and hence provide a key test of several dark matter models.
In this chapter, we described the process of inferring the properties of dark matter from strong lensing observations in terms of a Bayesian inference problem, and showed how it can be approached from different perspectives (Section \ref{sec:modelling}). For each possible prior choice and lens modelling technique, we then discussed their advantages and limitations (Section \ref{sec:modelling_approaches}). In Section \ref{sec:dsu} we provided an extensive discussion of the degeneracies and theoretical unknowns. In particular, we highlighted how better numerical simulations are needed to answer important and still open theoretical questions (Section \ref{sec:th_un}).

While we only focused on galaxy-scale lenses, all of the approaches discussed above can potentially be extended to cluster-scale lenses. These type of studies would allows one to probe larger cosmological volumes and hence potentially require fewer observations to reach an equal level of constraints on dark matter. However, the lensing potential in galaxy clusters is significantly more complex than those of galaxies and it remains yet to be demonstrated that one can constrain their mass distribution with enough precision to avoid contamination by complex macro-models (see Section \ref{sec:dsu_macro} for a discussion). 

Currently, constraints on dark matter from lensing are based on relative small samples of systems (about ten for compact sources and about 50 for extended ones). Ongoing and upcoming wide-sky surveys, with for example Euclid, the Vera Rubin Telescope, LOFAR and the SKA are expected to deliver orders of magnitudes more lensed galaxies and quasars. On the other hand, high-resolution follow-up with, for example, ALMA, the Global VLBI Network and the ELT will likely provide the data quality necessary to probe the dark matter distribution on small scales with extended sources as well as obtain better constraints on the macro-model for compact sources. We can expect, therefore, that in the next five to ten years, the field of strong gravitational lensing will provide robust and statistically meaningful constraints on the nature of dark matter.

\begin{acknowledgements}
S.~V. thanks the Max Planck Society for support through a Max Planck Lise Meitner Group and funding from the European Research Council (ERC) under the European Union's Horizon 2020 research and innovation programme (LEDA: grant agreement No 758853). G.~V. has received funding from the European Union's Horizon 2020 research and innovation programme under the Marie Sklodovska-Curie grant agreement No 897124. G.~D. was supported by a Gliese Fellowship. C.~D.~F acknowledges support for this work from the National Science Foundation under Grant No. AST-1715611.  D.~G. was partially supported by a HQP grant from the McDonald Institute (reference number HQP 2019-4-2). Y.~H. and L.~P. are supported by a generous donation by Eric and Wendy Schmidt with the recommendation of the Schmidt Futures Foundation, as well as the National Sciences and Engineering Council of Canada, the Fonds de recherche du Qu\'{e}bec, and the Canada Research Chairs Program. J.~P.~M. acknowledges support from the Netherlands Organization for Scientific Research (NWO, project number 629.001.023) and the Chinese Academy of Sciences (CAS, project number 114A11KYSB20170054).
\end{acknowledgements}

\bibliographystyle{aps-nameyear}      
\bibliography{references}                
\nocite{*}

\end{document}